%% file: main.tex
\documentclass[reprint,amsmath,amssymb,aps]{revtex4-2}
\usepackage{amssymb}
\usepackage{graphicx}
\usepackage{dcolumn}
\usepackage{bm}
\usepackage{color}
\usepackage{subfigure}
\usepackage{xspace}
\usepackage{textpos}
\usepackage[english]{babel}
\usepackage[switch]{lineno}
\usepackage{hyperref}

\begin{document}

\title{Measurement of the \texorpdfstring{$C\!P$}{CP}-even fraction of \texorpdfstring{$D^0\to K^+K^-\pi^+\pi^-$}{D2KKpipi} }

\begin{abstract}
    A determination of the $C\!P$-even fraction $F_+$ in the decay $D^0 \to K^+K^-\pi^+\pi^-$ is presented. Using $2.93$~fb$^{-1}$ of $e^+e^-\to\psi(3770)\to D\bar{D}$ data collected by the BESIII detector, one charm meson is reconstructed in the signal mode and the other in a $C\!P$ eigenstate or the decay $D\to K_{S, L}^0\pi^+\pi^-$. Analysis of the relative rates of these double-tagged events yields the result $F_+ = 0.730 \pm 0.037 \pm 0.021$, where the first uncertainty is statistical and the second is systematic. This is the first model-independent measurement of $F_+$ in $D^0 \to K^+K^-\pi^+\pi^-$ decays.
\end{abstract}

\date{13th December 2022}

\input{authorlist_2022-08-19}

\maketitle


\section{Introduction}
\label{section:Introduction}

\noindent The Standard Model description of $C\!P$ violation may be tested by measuring the lengths and angles of the Unitary Triangle of the CKM matrix \cite{Cabibbo:1963yz,Kobayashi:1973fv}. One of these angles, commonly denoted by $\gamma$\footnote{Also denoted $\phi_3$ in the literature}, is the only one accessible through tree-level processes, with negligible theoretical uncertainties~\cite{cite:BrodZupan}. Thus, a precise determination of $\gamma$ is an excellent Standard Model benchmark and direct measurements of $\gamma$ can be compared with indirect measurements that may be sensitive to new physics at loop level.

The angle $\gamma$ is conventionally measured in $B^\pm\to DK^\pm$ decays, where $D$ is a superposition of the flavor eigenstates $D^0$ and $\bar{D^0}$. An important class of $D$-meson decays for this purpose are those to $C\!P$ eigenstates~\cite{cite:GLW_paper}. Similarly, one can also use decay modes with mixed $C\!P$ content to measure $\gamma$, provided that this content is known~\cite{cite:Firstpipipi0,cite:pipipi0_CPfraction}. This content is parameterized by $F_+$, the $C\!P$-even fraction of the decay. Furthermore, decay modes with mixed $C\!P$ content can be used in studies of $D^0$-$\bar{D^0}$ oscillations, and searches for $C\!P$ violation in the charm system~\cite{SCMB_CPM_cite}.

This paper presents the first model-independent measurement of the $C\!P$-even fraction $F_+$ for the decay $D^0 \to K^+K^-\pi^+\pi^-$ using $2.93$~fb$^{-1}$ of quantum-correlated $\psi(3770)\to D\bar{D}$ data collected by the BESIII experiment. This measurement complements other strong-phase measurements performed with data collected by CLEO-c~\cite{cite:pipipi0_CPfraction, cite:cisi4pi, cite:KSpipipi0} and BESIII~\cite{cite:KSpipiStrongPhase, cite:cisiKSKK, cite:K3piStrongPhase,BESIII:fourpi}, and it is an important input to future analyses of $\gamma$ and $D^0$-$\bar{D^0}$ oscillations using this channel.

\section{Measurement strategy}
\label{section:Measurement_strategy}

\noindent The strong decay of $\psi(3770)\to D\bar{D}$ conserves the $C = -1$ quantum number of the initial state, leaving the $D$-meson pair in an anti-symmetric wave function. This quantum correlation allows for a direct access to the strong-phase difference between $D^0$ and $\bar{D^0}$ decays through a double-tag (DT) analysis. The method uses single-tag (ST) events, which are events where one of the charm mesons is reconstructed in a $C\!P$ eigenstate, with no requirements on the decay of the other meson, and DT events, where both $D$ mesons are reconstructed, one in a tag mode and the other in the signal mode.

Table~\ref{table:Tag_modes} lists all the tag modes used for this analysis. The analysis can be split into three categories: $C\!P$ tags, $K_S^0\pi^+\pi^-$, and $K_L^0\pi^+\pi^-$. The $C\!P$ tags are modes in which the $D$ meson decays to a $C\!P$ eigenstate. The modes $D\to K_{S, L}^0\pi^+\pi^-$ are of mixed $C\!P$ content, since these decays can proceed through both $C\!P$-even and $C\!P$-odd amplitudes. The mode $\pi^+\pi^-\pi^0$ is listed as a $C\!P$-even tag since its $C\!P$-even fraction, $F_+^{\pi\pi\pi^0} = 0.973 \pm 0.017$~\cite{cite:pipipi0_CPfraction}, is very close to unity. The modes $D\to K_L^0\pi^0\pi^0$, $D\to K_L^0\omega$, and the self-tag $D\to K^+K^-\pi^+\pi^-$ have not been included because their yields are low and the inclusion of these tag modes would not significantly improve the precision of the measurement.

\begin{table}[htb]
    \centering
    \caption{Tag modes used in this analysis.}
    \label{table:Tag_modes}
    \begin{tabular}{cc}
        \hline
        Category     & Tag modes \\
        \hline
        $C\!P$ even  & $K^+K^-$, $\pi^+\pi^-$, $K_S^0\pi^0\pi^0$, $\pi^+\pi^-\pi^0$, $K_L^0\pi^0$ \\
        $C\!P$ odd   & $K_S^0\pi^0$, $K_S^0\eta_{\gamma\gamma}$, $K_S^0\eta^\prime(\pi\pi\eta)$, $K_S^0\eta^\prime(\rho^0\gamma)$, $K_S^0\omega$ \\
        Mixed $C\!P$ & $K_S^0\pi^+\pi^-$, $K_L^0\pi^+\pi^-$ \\
        \hline
    \end{tabular}
\end{table}

The predicted ST yield of a tag mode $D\to f$ with $C\!P$-even fraction $F_+^f$ is given by
\begin{equation}
    N^{\rm ST}(f) = 2N_{D\bar{D}}\mathcal{B}(f)\epsilon_{\rm ST}(f)\big(1 - (2F_+^{f} - 1)y\big),
    \label{equation:ST_yield}
\end{equation}
where $N_{D\bar{D}} = (10597 \pm 28 \pm 98) \times 10^3$~\cite{cite:NDD} is the total number of $D\bar{D}$ pairs, $\mathcal{B}$ is the branching fraction, $\epsilon_{\rm ST}$ is the reconstruction efficiency of the ST mode, and $y = (0.615^{+0.056}_{-0.055})\times 10^{-2}$ is the charm-mixing parameter~\cite{cite:HFLAV2018}. In Eq.~\eqref{equation:ST_yield} and subsequent expressions, $\mathcal{O}(y^2)$ terms are neglected. For pure $C\!P$-even (odd) tags, $F_+^f = 1$ ($0$).

Events where one $D$ meson is reconstructed as the signal decay $D\to K^+K^-\pi^+\pi^-$, while the other is reconstructed as a pure or mixed-$C\!P$ tag mode $f$, have a predicted DT yield
\begin{linenomath}
    \begin{align}
        N^{\rm DT}(KK\pi\pi | f) =& 2N_{D\bar{D}}\mathcal{B}(f)\mathcal{B}(KK\pi\pi)\epsilon_{\rm DT}(KK\pi\pi | f) \nonumber \\
        \times&\big(1 - (2F_+^{f} - 1)(2F_+ - 1)\big),
        \label{equation:DT_yield}
    \end{align}
\end{linenomath}
where $\mathcal{B}(KK\pi\pi)$ is the branching fraction of $D^0\to K^+K^-\pi^+\pi^-$, $\epsilon_{\rm DT}$ is the reconstruction efficiency of the DT event, and $F_+$ denotes the $C\!P$-even fraction of $D^0\to K^+K^-\pi^+\pi^-$. Equations~\eqref{equation:ST_yield} and~\eqref{equation:DT_yield} can be combined into
\begin{linenomath}
    \begin{align}
        \frac{N^{\rm DT}(KK\pi\pi | f)}{N^{\rm ST}(f)/\big(1 - (2F_+^{f} - 1)y\big)}\times\frac{\epsilon_{\rm ST}(f)}{\epsilon_{\rm DT}(KK\pi\pi | f)} =& \nonumber \\
        \mathcal{B}(KK\pi\pi)\big(1 - (2F_+^{f} - 1)(2F_+ - 1)\big).
        \label{equation:DT_ST_yield_ratio}
    \end{align}
\end{linenomath}
Equation~\eqref{equation:DT_ST_yield_ratio} indicates that the ratio of the DT to ST yields, after efficiency corrections, is sensitive to $\mathcal{B}(KK\pi\pi)$ and the $C\!P$-even fraction $F_+$. Measuring this quantity for tags of different $C\!P$ eigenvalues allows $F_+$ to be determined.

For the $K_S^0\pi^+\pi^-$ tag, which is a decay mode of  mixed $C\!P$, an enhanced sensitivity to $F_+$ is obtained by separating events into bins of phase space of the tag decay. The amplitude-averaged strong-phase difference between $D^0$ and $\bar{D^0}$ decays has been measured in these bins by both CLEO~\cite{cite:CLEOcisiKSpipi} and BESIII~\cite{cite:KSpipiStrongPhase}. The binning scheme used for this analysis is the ``equal $\Delta\delta_D$ binning", where bin boundaries are chosen such that each bin spans an equal range in the strong-phase difference. The fractional yield $K_i$ of $D^0$ decays and amplitude-averaged cosine of the strong-phase difference $c_i$ have been measured in each bin. There are eight pairs of bins in total~\cite{cite:CLEOcisiKSpipi}, and since each pair of bins have the same value for $c_i$, the data in each pair are merged. The combined CLEO and BESIII results from Ref.~\cite{cite:KSpipiStrongPhase} are used, and they are treated as external inputs to the $F_+$ determination. The yield-ratio expression in bin $i$ is
\begin{linenomath}
    \begin{align}
        \frac{N_i^{\rm DT}(KK\pi\pi | f)}{N^{\rm ST}(f)/\big(1 - (2F_+^{f} - 1)y\big)}\times\frac{\epsilon_{\rm ST}(f)}{\epsilon_{\rm DT}(KK\pi\pi | f)} = \nonumber \\
        \mathcal{B}(KK\pi\pi)\big(K_i + K_{-i} - 2\sqrt{K_iK_{-i}}c_i(2F_+ - 1)\big).
        \label{equation:DT_ST_yield_ratio_binned}
    \end{align}
\end{linenomath}
The expression for the $D\to K_L^0\pi^+\pi^-$ tag is obtained by replacing $K_i$ with $K_i^\prime$ and $c_i$ with $-c_i^\prime$, the values of which are also reported in Refs.~\cite{cite:CLEOcisiKSpipi,cite:KSpipiStrongPhase}.

\section{BEPCII and the BESIII detector}
\label{section:Description_of_the_BEPCII_and_the_BESIII_detector}

\noindent The BESIII detector~\cite{Ablikim:2009aa} records symmetric $e^+e^-$ collisions provided by the BEPCII storage ring~\cite{Yu:2016cof}, which operates with a centre-of-mass energy range from $\sqrt{s} = 2.00$~GeV to $4.95$~GeV, with a peak luminosity of $1\times10^{33}$~cm$^{-2}$s$^{-1}$ achieved at $\sqrt{s} = 3.773$~GeV. \mbox{BESIII} has collected large data samples in this energy region~\cite{Ablikim:2019hff}. The cylindrical core of the BESIII detector covers 93\% of the full solid angle and consists of a helium-based multilayer drift chamber~(MDC), a plastic scintillator time-of-flight system~(TOF), and a CsI(Tl) electromagnetic calorimeter~(EMC), which are all enclosed in a superconducting solenoidal magnet providing a 1.0~T magnetic field. The solenoid is supported by an octagonal flux-return yoke with resistive plate counter muon-identification modules interleaved with steel. The charged-particle momentum resolution at $1~{\rm GeV}/c$ is $0.5\%$, and the resolution of the rate of energy loss, ${\rm d}E/{\rm d}x$, is $6\%$ for electrons from Bhabha scattering. The EMC measures photon energies with a resolution of $2.5\%$ ($5\%$) at $1$~GeV in the barrel (end-cap) region. The time resolution in the TOF barrel region is 68~ps, while that in the end-cap region is 110~ps.

Simulated data samples produced with a {\sc geant4}-based~\cite{geant4} Monte Carlo (MC) package, which includes the geometric description of the BESIII detector and the detector response, are used to determine detection efficiencies and to estimate backgrounds. The simulation models the beam-energy spread and initial-state radiation in the $e^+e^-$ annihilations with the generator {\sc kkmc}~\cite{ref:kkmc}. The inclusive MC sample includes the production of $D\bar{D}$ pairs, the non-$D\bar{D}$ decays of the $\psi(3770)$, the initial-state radiation production of the $J/\psi$ and $\psi(3686)$ states, and the continuum processes incorporated in {\sc kkmc}~\cite{ref:kkmc}. All particle decays are modelled with {\sc evtgen}~\cite{Lange:2001uf,Ping:2008zz} using branching fractions either taken from the Particle Data Group~\cite{pdg}, when available, or otherwise estimated with {\sc lundcharm}~\cite{YANGRui-Ling:61301,Chen:2000tv}. Final-state radiation from charged final-state particles is incorporated using the {\sc photos} package~\cite{RICHTERWAS1993163}.

To ensure the best possible description of the distribution of the $D \to K^+K^-\pi^+\pi^-$ decays in phase space, the simulation samples are reweighted using the most recent amplitude model for this decay~\cite{LHCb-PAPER-2018-041}. Quantum correlations are accounted for in the reweighting.

\section{Event selection}
\label{section:Event_selection}

\noindent Charged tracks detected in the MDC are required to be within a polar angle ($\theta$) range of $|\!\cos\theta|<0.93$, where $\theta$ is defined with respect to the $z$-axis, which is the symmetry axis of the MDC. For charged tracks not originating from $K_S^0$ decays, the distance of closest approach to the interaction point (IP), $|V_{z}|$, must be less than 10\,cm along the $z$-axis, and less than 1\,cm in the transverse plane.

Photon candidates are identified using showers in the EMC. The deposited energy of each shower must be more than 25~MeV in the barrel region ($|\cos \theta|< 0.80$) and more than 50~MeV in the end-cap region ($0.86 <|\!\cos \theta|< 0.92$). To exclude showers that originate from charged tracks, the angle subtended by the EMC shower and the position of the closest charged track at the EMC must be greater than 10 degrees as measured from the IP. To suppress electronic noise and showers unrelated to the event, the difference between the EMC time and the event-start time is required to be within [0, 700]\,ns.

Particle identification~(PID) for charged tracks combines measurements of the d$E$/d$x$ in the MDC, and the time of flight as measured by the TOF system, to form likelihoods $\mathcal{L}(h)$ for each hadron hypothesis $h$ $(h=K,\pi)$. Charged kaons and pions are identified by comparing the likelihoods for the kaon and pion hypotheses, $\mathcal{L}(K)>\mathcal{L}(\pi)$ and $\mathcal{L}(\pi)>\mathcal{L}(K)$, respectively.

Each $K_{S}^0$ candidate is reconstructed from two oppositely charged tracks satisfying $|V_{z}|<$ 20~cm. The two charged tracks are assigned the pion hypothesis without imposing PID criteria. They are constrained to originate from a common vertex and are required to have an invariant mass within 12~MeV$/c^{2}$ of the known $K^0_{S}$ mass~\cite{pdg}. The decay length of the $K^0_S$ candidate is required to be greater than twice the vertex resolution away from the IP.

Candidate $\pi^0$ and $\eta$ mesons are reconstructed through the decays $\pi^0\to\gamma\gamma$ and $\eta\to\gamma\gamma$, with their di-photon invariant masses required to be within $[115, 150]$ and $[480, 580]$~MeV$/c^2$, respectively. The $\eta^\prime$ meson is reconstructed through  $\eta^\prime\to\pi^+\pi^-\eta$ and $\rho^0(\pi^+\pi^-)\gamma$, with the invariant masses of the decay products within $[940, 976]$ and $[940, 970]$~MeV$/c^2$. The invariant mass of the pion pair in the $\rho^0$ decay must lie within $[626, 924]$~MeV$/c^2$.

In the reconstruction of $D\to\pi^+\pi^-\pi^0$ and $D\to K^+K^-\pi^+\pi^-$ decays, the $\pi^+\pi^-$ pair is required to originate from a vertex within twice the vertex resolution from the IP, in order to reduce backgrounds from $D\to K_S^0\pi^0$ and $D\to K_S^0K^+K^-$, respectively. For $\pi^+\pi^-\pi^0$, the $\pi^+\pi^-$ invariant mass is required to be more than $18$~MeV$/c^2$ away from the known $K_S^0$ mass. For $K^+K^-\pi^+\pi^-$, the $\pi^+\pi^-$ mass must fall outside $[477, 507]$~MeV$/c^2$.

For fully reconstructed tags, $\Delta E = E_D - \sqrt{s}/2$, where $E_D$ is the reconstructed energy of the $D$ meson, is required to be within $3\sigma$ of the signal peak. This requirement removes combinatorial background. In the tags containing a $K_L^0$, a partial reconstruction is performed where the signal $D\to K^+K^-\pi^+\pi^-$ is first reconstructed. From the remaining tracks and showers, the tag mode is reconstructed without the $K_L^0$ meson. It is required that there are no additional charged tracks or $\pi^0$ candidates. The $K_L^0$ momentum is then inferred from the missing momentum of the event. Since there is a missing particle, the ST yield cannot be measured in tag modes containing a $K_L^0$ meson.

Additionally, to increase the yield of $K^+K^-\pi^+\pi^-$ vs $K_S^0\pi^+\pi^-$ DTs, events where a charged kaon is not reconstructed are also considered. The tag mode $D\to K_S^0\pi^+\pi^-$ is first reconstructed, and it is required that there are exactly three remaining tracks, identified as a kaon and two oppositely charged pions. The momentum of the charged kaon that is not reconstructed is inferred from the missing momentum. To reduce the background from $D\to K^-\pi^+\pi^-\pi^+\pi^0$ decays, it is required that there are no $\pi^0$ candidates in the event.

In the $K_{S, L}^0\pi^+\pi^-$ tags, a Kalman kinematic fit~\cite{KalmanFit_cite} is performed to improve the resolution of the final-state particle momenta by constraining the $K_{S, L}^0$ and $D$ invariant masses to their known values~\cite{pdg}.

\section{Single- and double-tag yield determination}
\label{section:Single_and_double_tag_yield_determination}

\begin{figure*}[htb]
    \centering
    \includegraphics[height=3.55cm,trim={-40.0cm 20.0cm -38.3cm 1.5cm},clip]{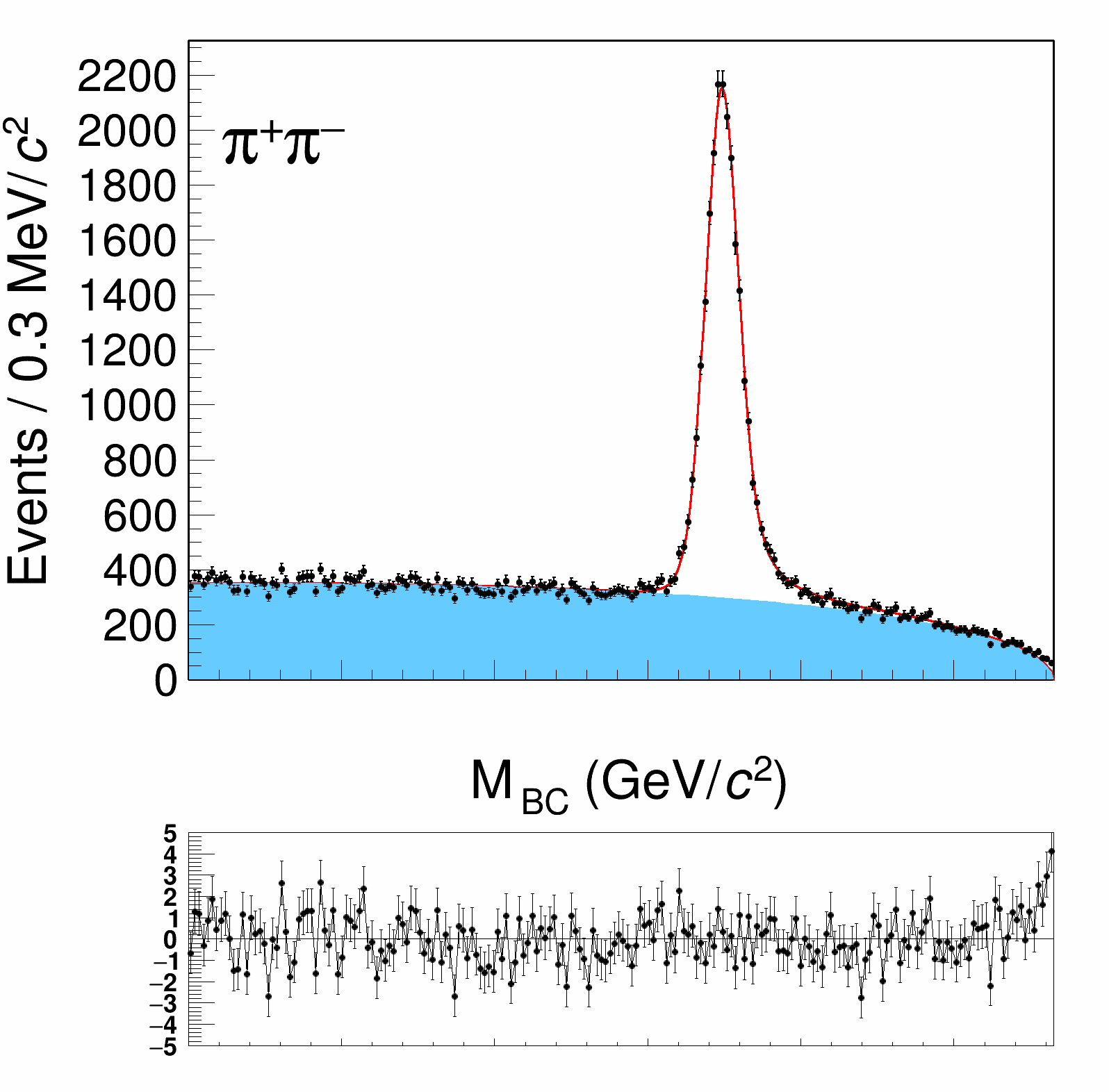}
    \includegraphics[height=3.55cm,trim={0.0cm 20.0cm 2.5cm 1.5cm},clip]{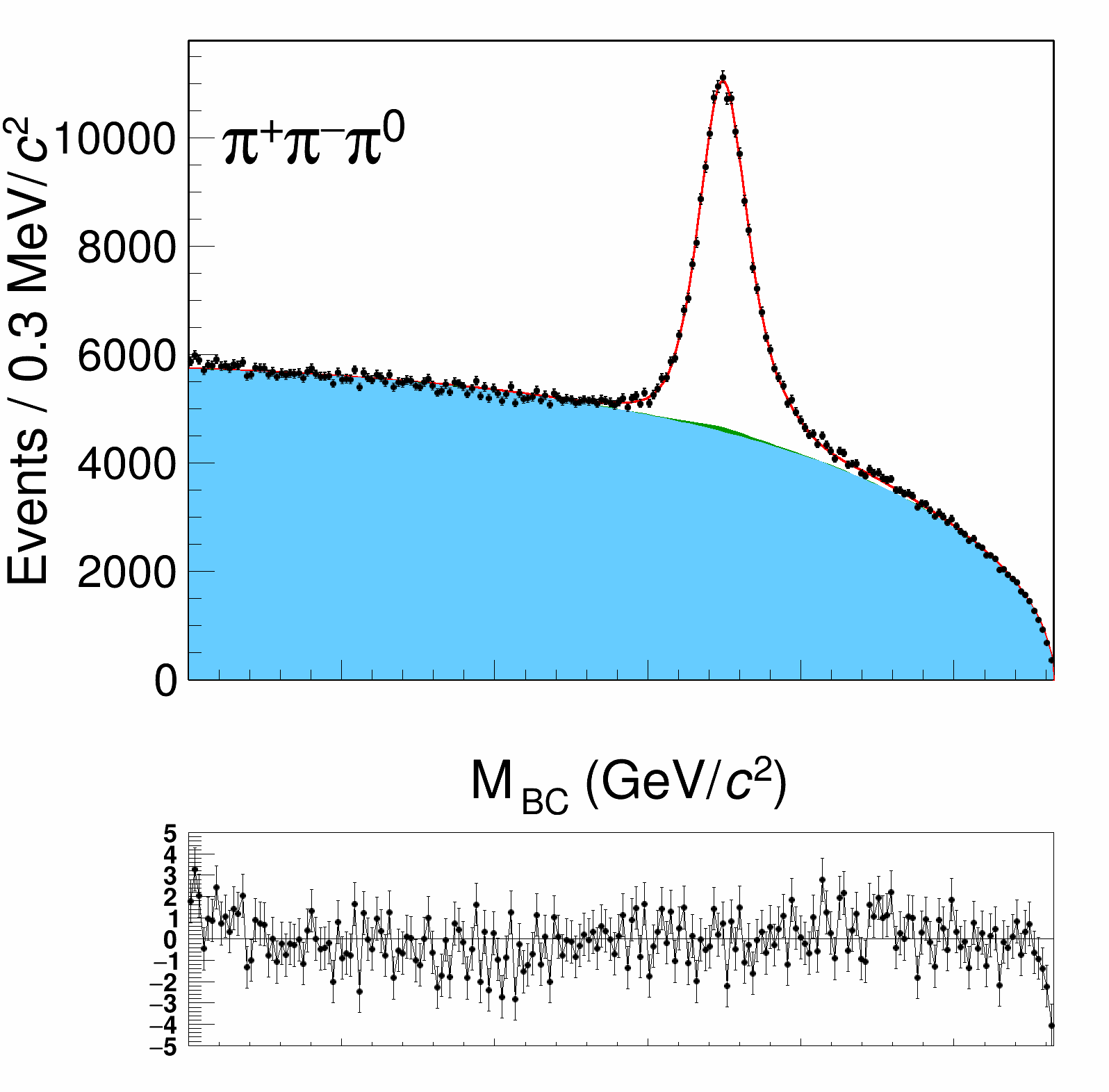}
    \includegraphics[height=3.55cm,trim={4.0cm 20.0cm 2.5cm 1.5cm},clip]{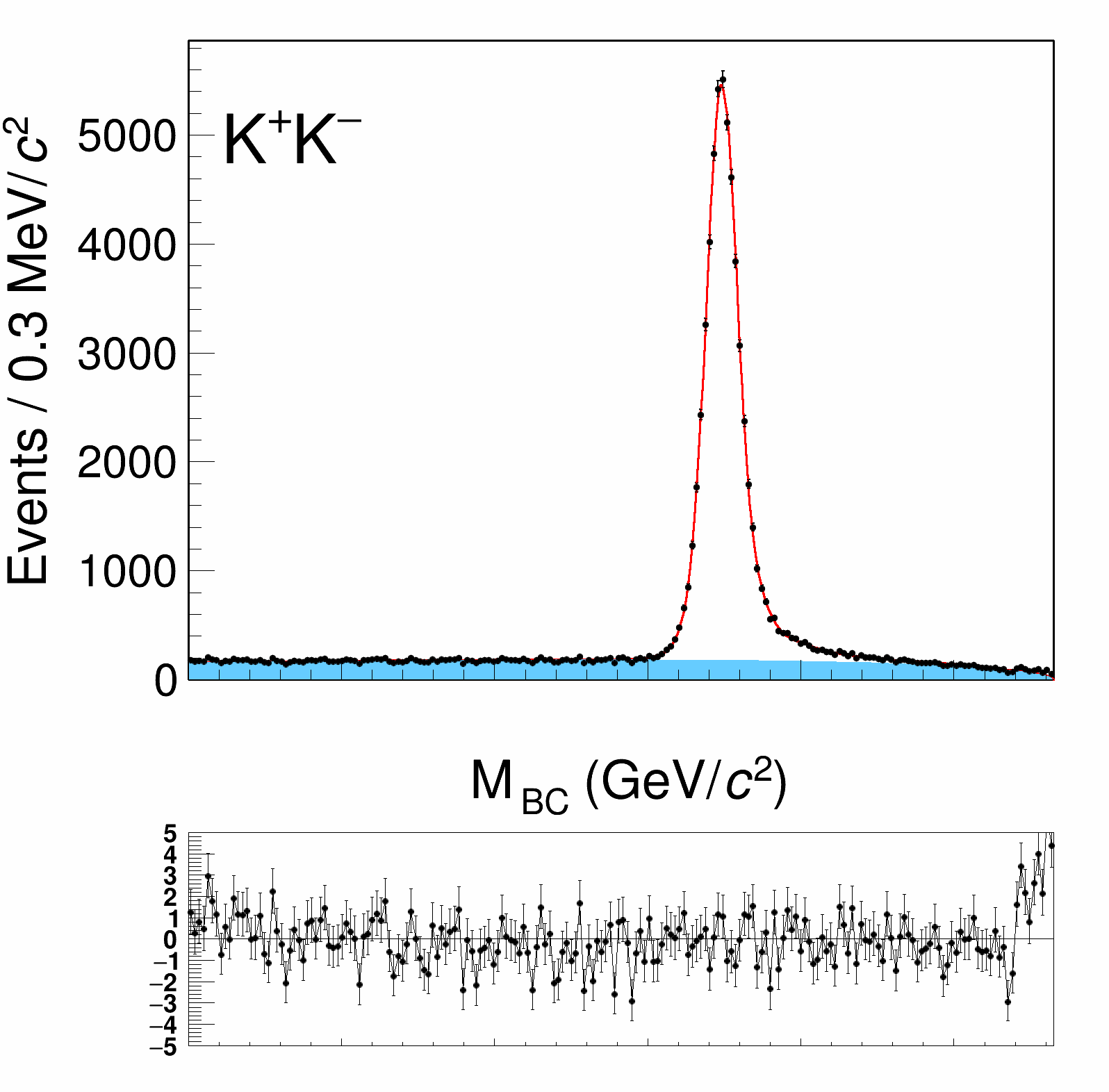}
    \includegraphics[height=3.55cm,trim={4.0cm 20.0cm 2.5cm 1.5cm},clip]{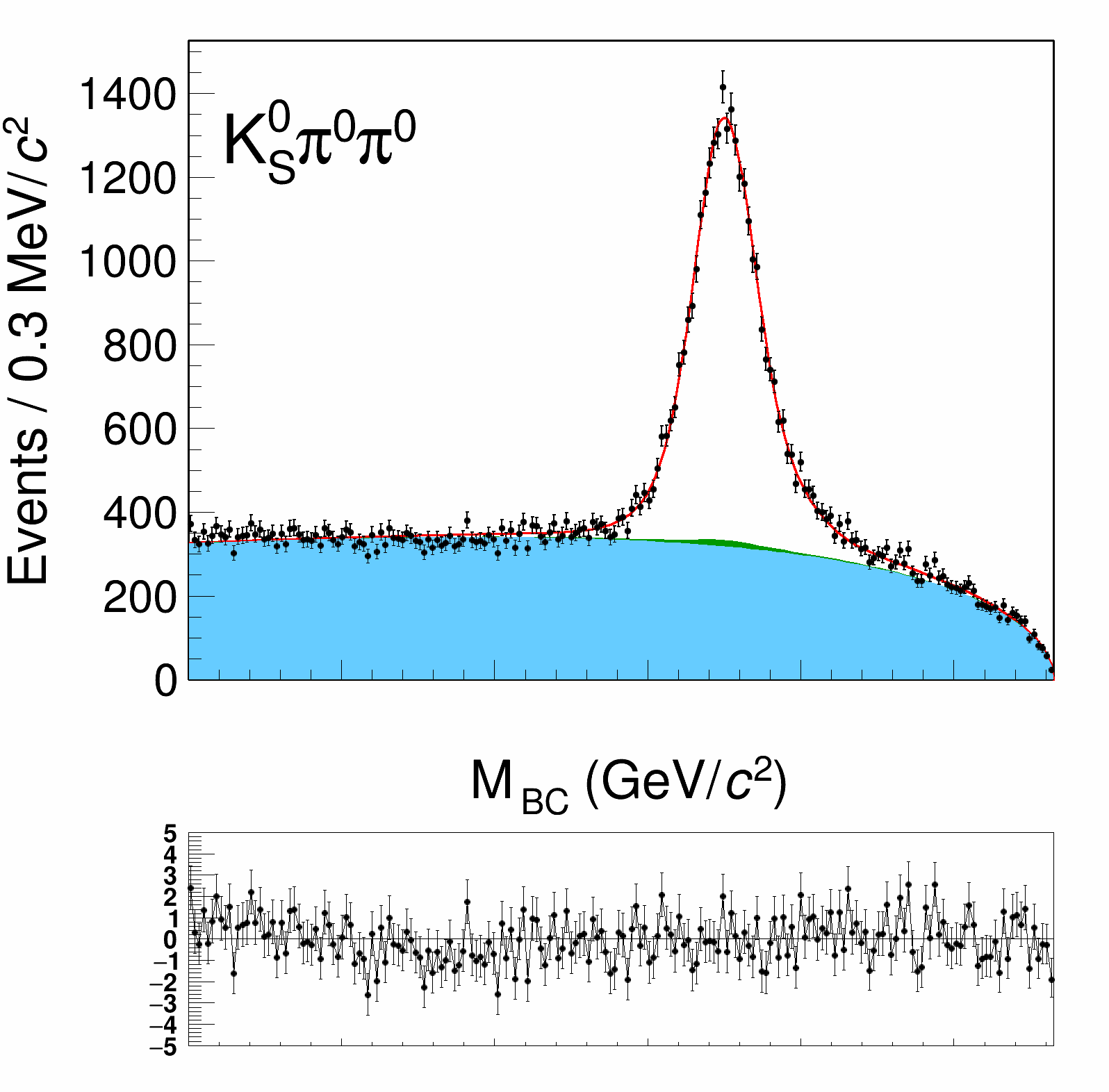}
    \includegraphics[height=3.55cm,trim={0.0cm 20.0cm 2.5cm 1.5cm},clip]{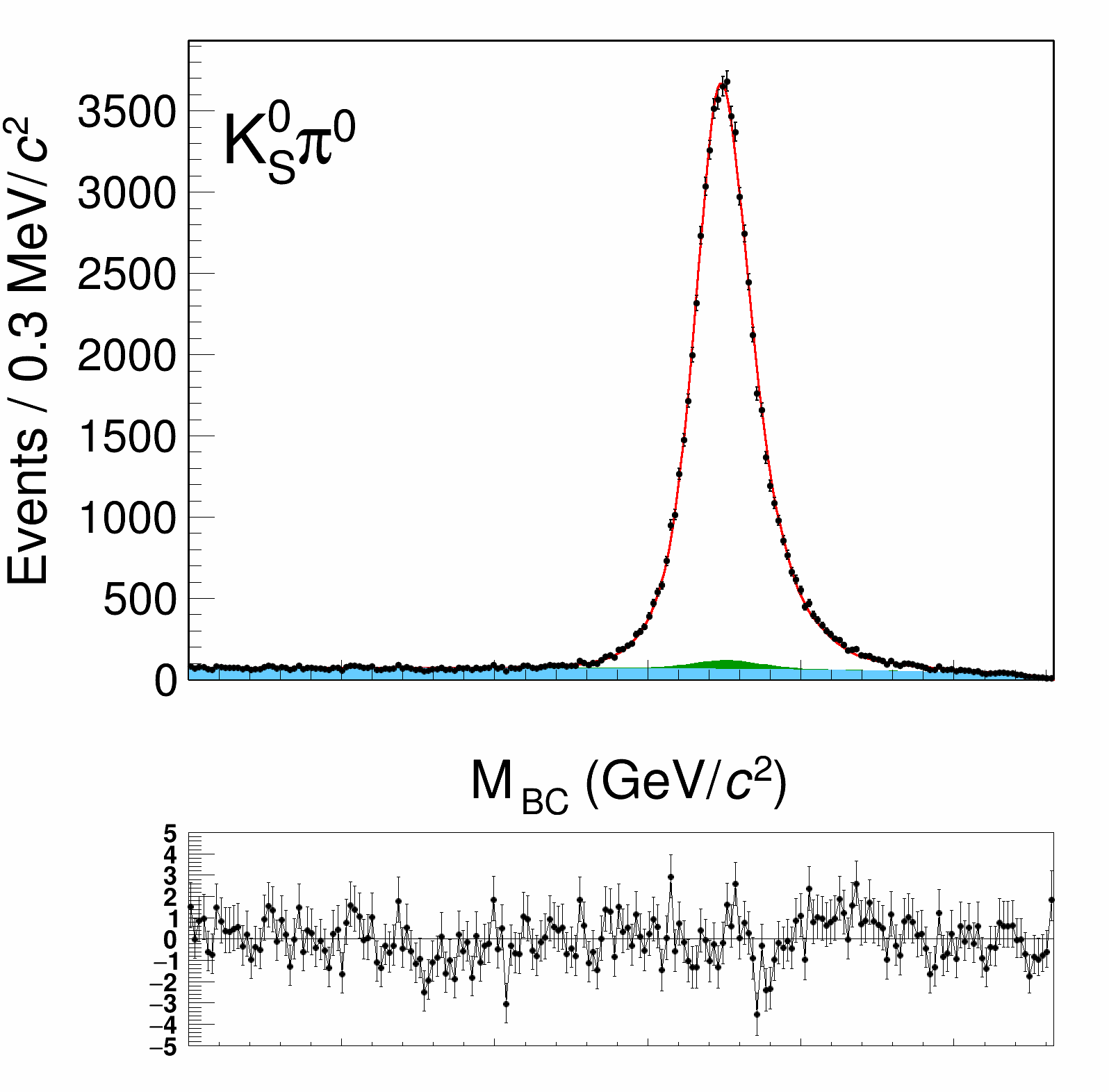}
    \includegraphics[height=3.55cm,trim={4.0cm 20.0cm 2.5cm 1.5cm},clip]{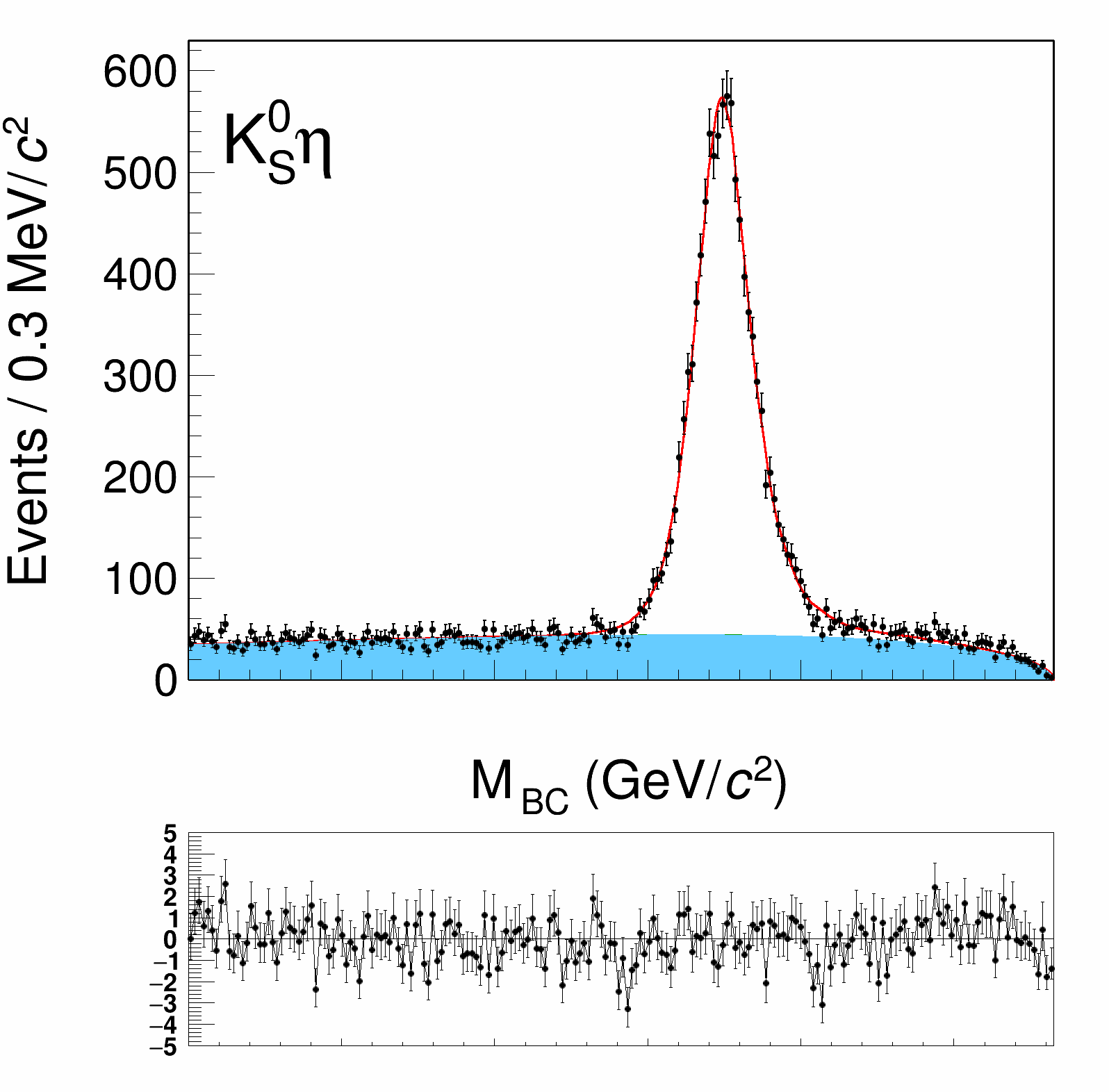}
    \includegraphics[height=3.55cm,trim={4.0cm 20.0cm 2.5cm 1.5cm},clip]{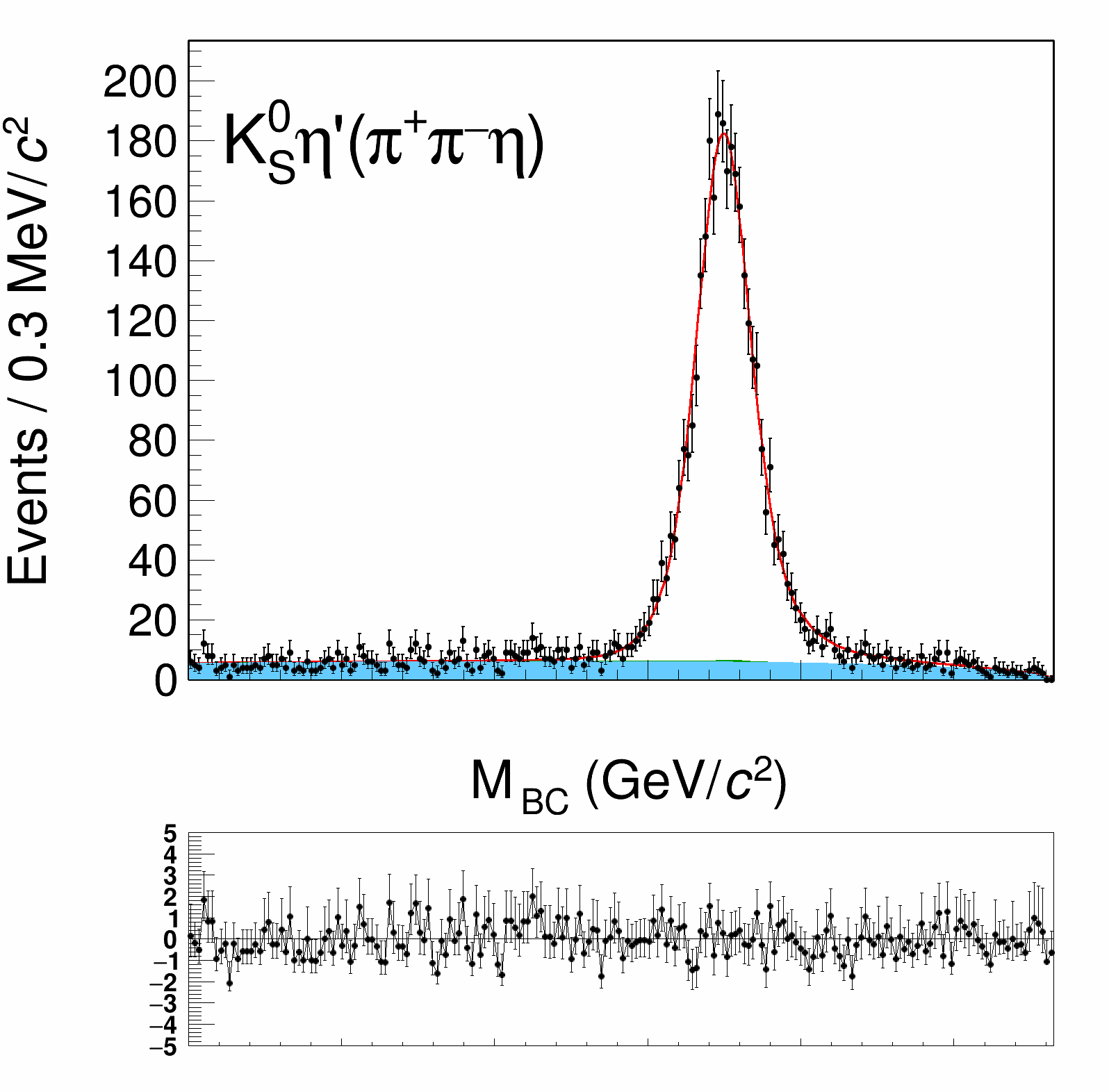}
    \includegraphics[height=4.2cm,trim={0.0cm 13.8cm 2.5cm 1.5cm},clip]{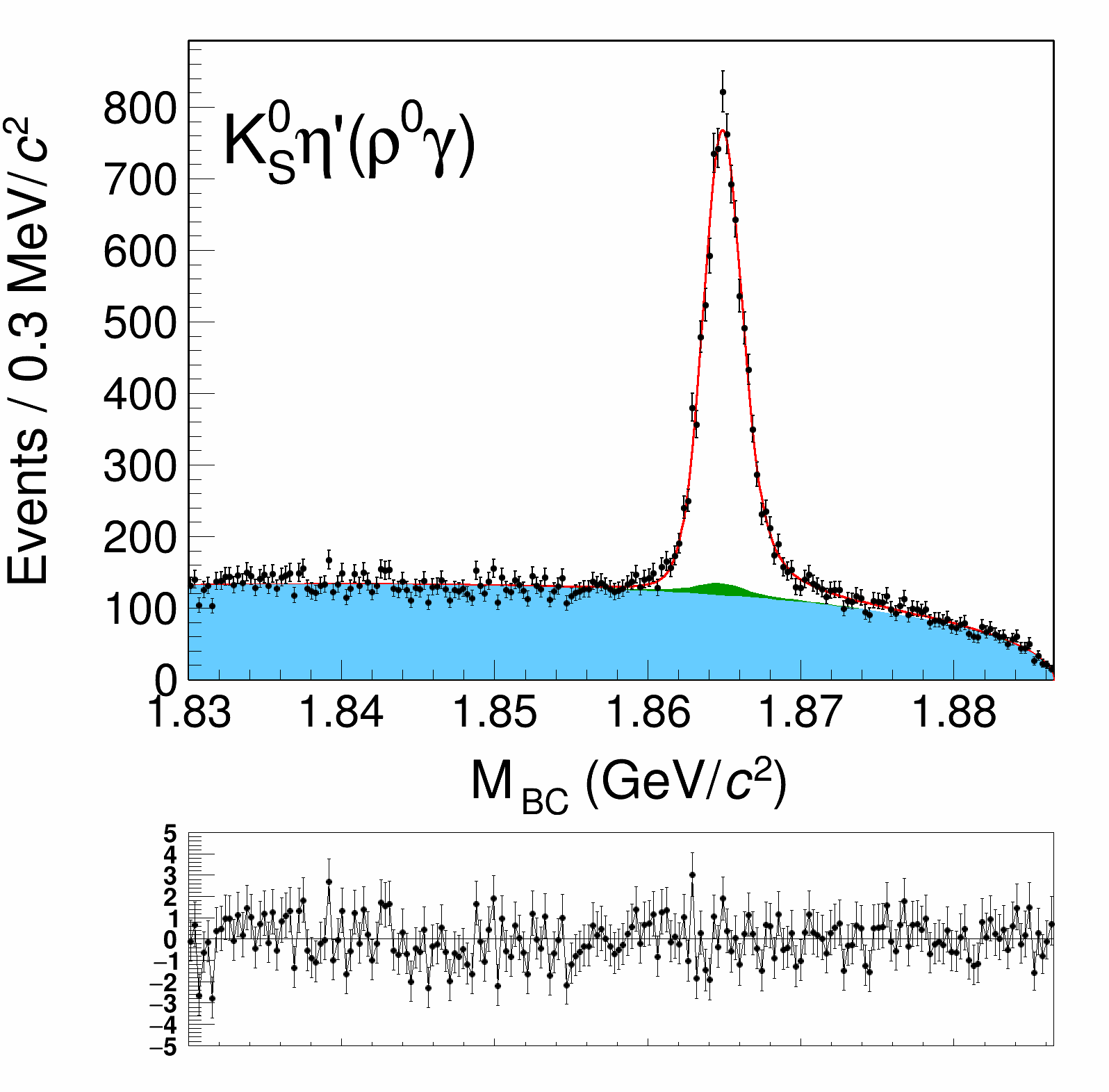}
    \includegraphics[height=4.2cm,trim={4.0cm 13.8cm 2.5cm 1.5cm},clip]{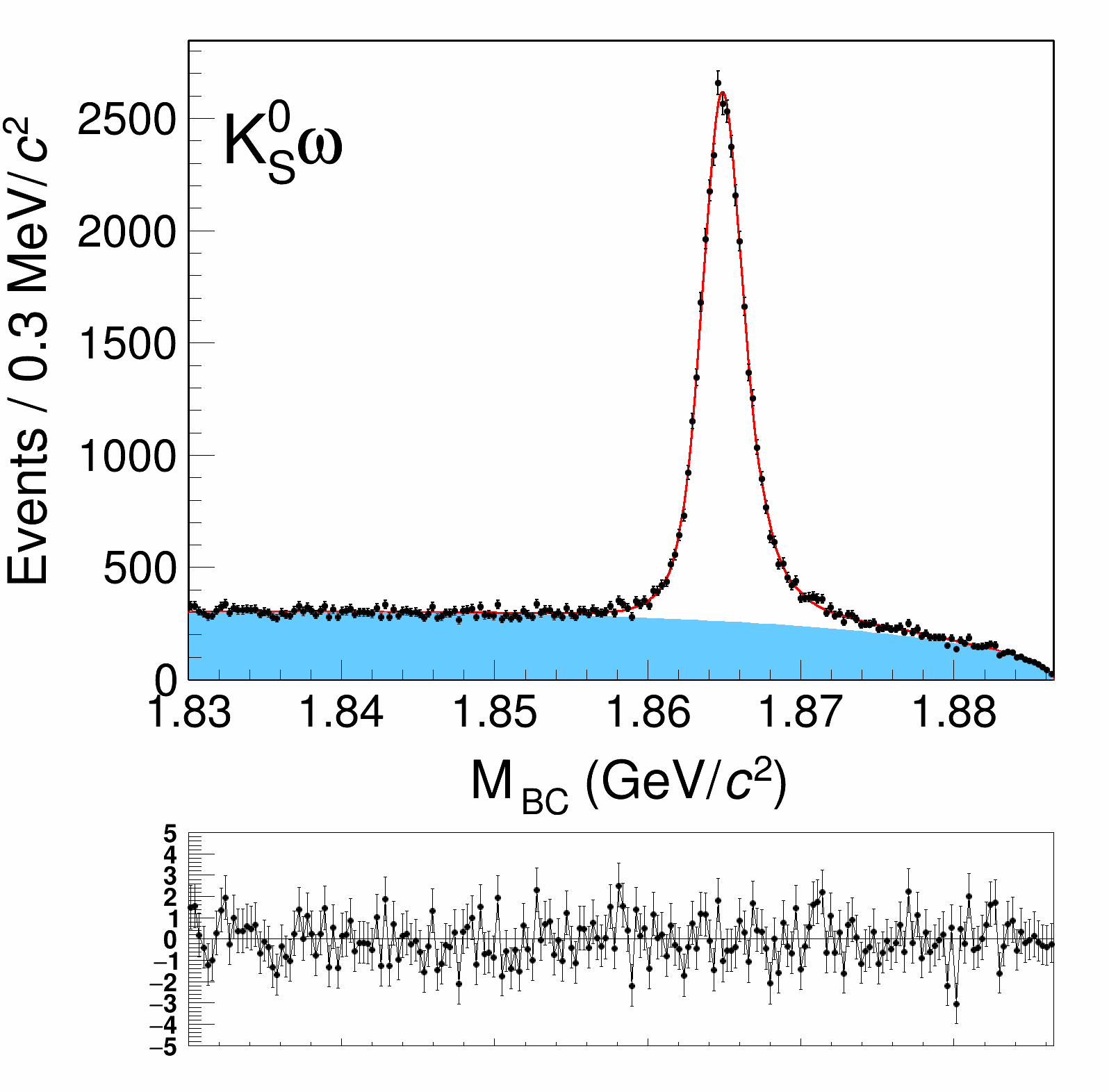}
    \includegraphics[height=4.2cm,trim={4.0cm 13.8cm 2.5cm 1.5cm},clip]{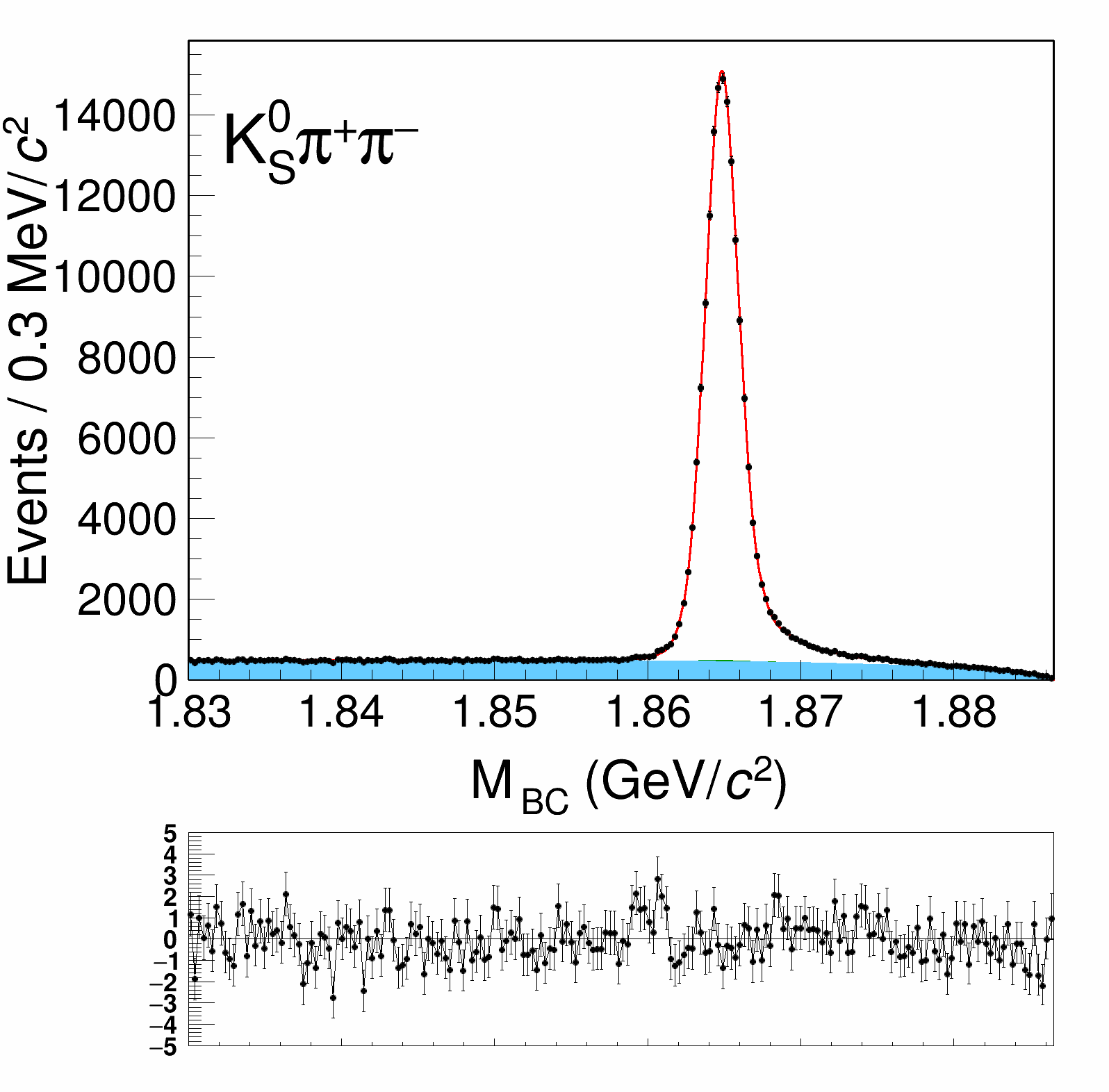}
    \caption{ST $M_{\rm BC}$ distributions. Data points are shown in black with error bars and the red curve is the fit result. The solid blue shape is combinatorial background. The green, stacked on top of the blue, is peaking background.}
    \label{figure:ST_MBC}
\end{figure*}

\noindent The ST yield of each fully reconstructed tag mode is determined by a maximum-likelihood fit of the beam-constrained mass $M_{\rm BC} = \sqrt{E_{\rm beam}^2 - \big\lvert\sum_i\vec{p}_i\big\rvert^2}$, where the sum runs over the momenta $\vec{p}_i$ of all the $D$ decay products. The signal shape is obtained from simulation, but convolved with a Gaussian function to account for differences in resolution between data and simulation. The width of the Gaussian function is a free parameter in the fit and the difference in resolution between data and simulation was found to be a few hundred ${\rm keV}/c^2$. Since the difference is small, no further correction to the simulation is performed. The combinatorial background is modelled by an ARGUS function~\cite{cite:Argus}. The $M_{\rm BC}$ distributions and the fitted shapes are shown in Fig.~\ref{figure:ST_MBC}. In all cases the fit quality is found to be good.

For the partially reconstructed tag mode $D\to K_L^0\pi^0$, the ST yield cannot be measured directly. Nonetheless, an effective ST reconstruction efficiency is calculated from $\epsilon_{\rm ST}(K_L\pi^0) = \epsilon_{\rm DT}(KK\pi\pi|K_L\pi^0)/\epsilon_{\rm ST}(KK\pi\pi)$. The effective ST yield is then calculated from this efficiency, the branching fraction~\cite{cite:deltaKpi}, and $N_{D\bar{D}}$, using Eq.~\eqref{equation:ST_yield}. The ST yields and their efficiencies, determined from simulation, are presented in Table~\ref{table:Single_tag_yields_efficiencies}. The ST yields are in good agreement with the results from Refs.~\cite{cite:KSpipiStrongPhase, cite:cisiKSKK, BESIII:fourpi}.

The level of peaking background in the fully reconstructed tag modes is around $1\%$ or less. In tag modes containing a $K_L^0$ there is a larger contamination from $K_S^0\to\pi^0\pi^0$ decays, where the $\pi^0$ mesons are not reconstructed. This peaking background is found from simulation to be around $6\%$ of the signal yield. The shapes of peaking backgrounds are fixed from simulation samples, while the yields are calculated from the branching fractions and efficiencies, relative to that of the signal yield.

\begin{table}[htb]
    \centering
    \caption{ST yields and efficiencies for $C\!P$ tags.  In the case of modes involving  $K_L^0$ mesons, these are effective quantities, as explained in the text. The uncertainties are statistical only.} \vspace{0.2cm}
    \label{table:Single_tag_yields_efficiencies}
    \begin{tabular}{lcc}
        \hline
        Tag                                & Yield                & Efficiency ($\%$) \\
        \hline
        $K_S^0\omega$                      & $22068 \pm 217$      & $14.50 \pm 0.08$     \\
        $K_S^0\eta^\prime(\pi^+\pi^-\eta)$ & \phantom{0}$3213 \pm 62$\phantom{0}        & $12.81 \pm 0.07$     \\
        $K_S^0\eta^\prime(\rho^0\gamma)$   & \phantom{0}$8283 \pm 116$       & $20.80 \pm 0.09$     \\
        $K_S^0\eta$                        & \phantom{0}$9308 \pm 113$       & $31.78 \pm 0.10$     \\
        $K_S^0\pi^0$                       & $67876 \pm 278$      & $38.18 \pm 0.11$     \\
        $\pi^+\pi^-\pi^0$                  & $107504 \pm 602$\phantom{0}     & $36.65 \pm 0.11$     \\
        $\pi^+\pi^-$                       & $20386 \pm 179$      & $67.41 \pm 0.10$     \\
        $K_S^0\pi^0\pi^0$                  & $22392 \pm 229$      & $14.35 \pm 0.08$     \\
        $K_L^0\pi^0$                       & \phantom{0}$47595 \pm 1653$     & $27.83 \pm 0.23$     \\
        $K^+K^-$                           & $56303 \pm 262$      & $63.41 \pm 0.11$     \\
        \hline
    \end{tabular}
\end{table}

Similarly, for fully reconstructed DT events, the $M_{\rm BC}$ on the signal side is fitted. The approach is identical to that for ST candidates, but corrections are applied to the peaking-background estimates to account for enhancements and suppressions due to quantum correlations. The quantum-correlation corrections are calculated using knowledge of the $C\!P$ contents of both the signal and tag modes. The $C\!P$ content of $D^0\to K^+K^-\pi^+\pi^-$ is obtained from the amplitude model in Ref.~\cite{LHCb-PAPER-2018-041}.

\begin{figure*}[htb]
    \centering
    \includegraphics[height=3.6cm,trim={-40.0cm 20.0cm -38.3cm 1.5cm},clip]{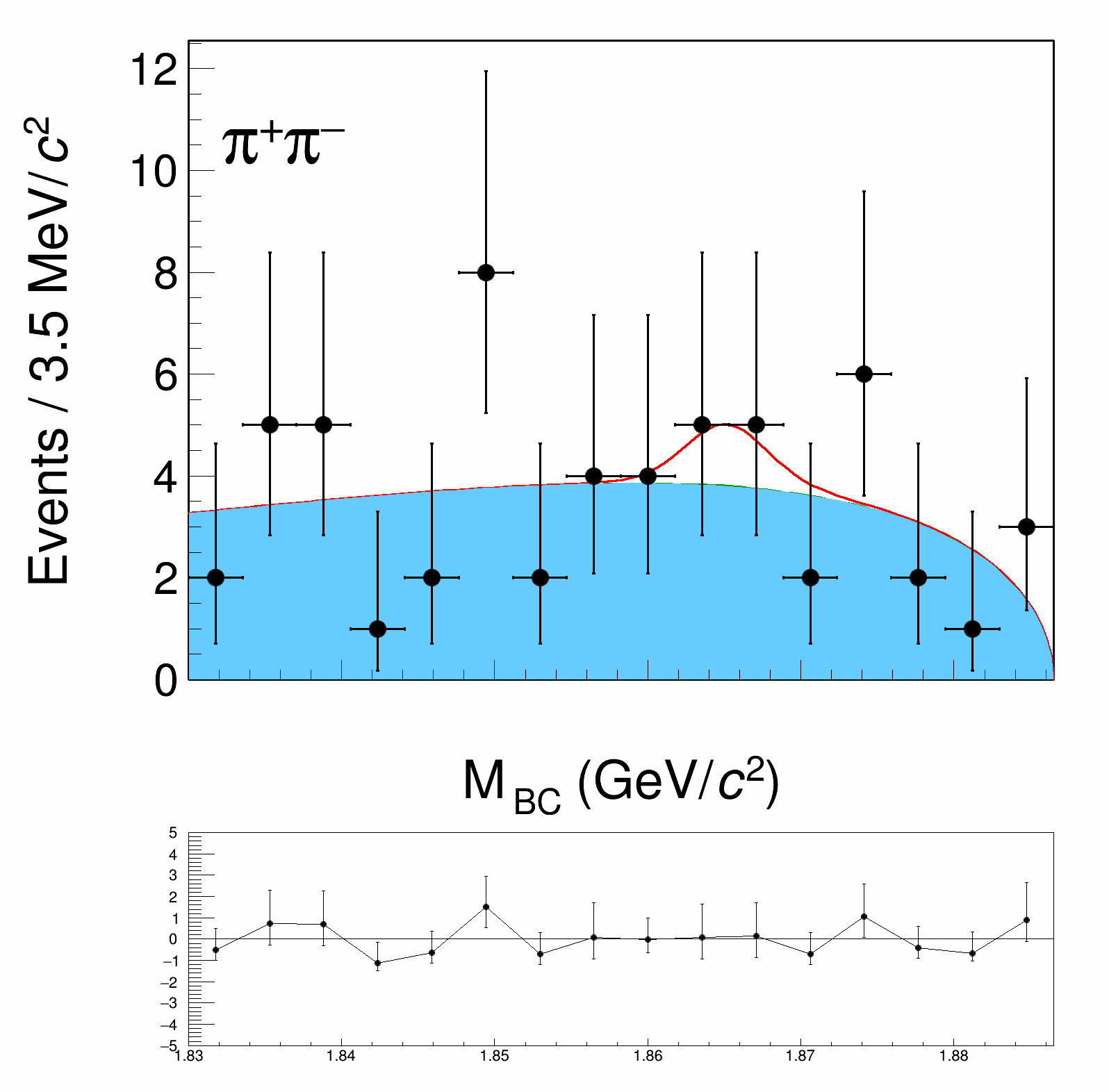}
    \includegraphics[height=3.6cm,trim={1.0cm 20.0cm 2.5cm 1.5cm},clip]{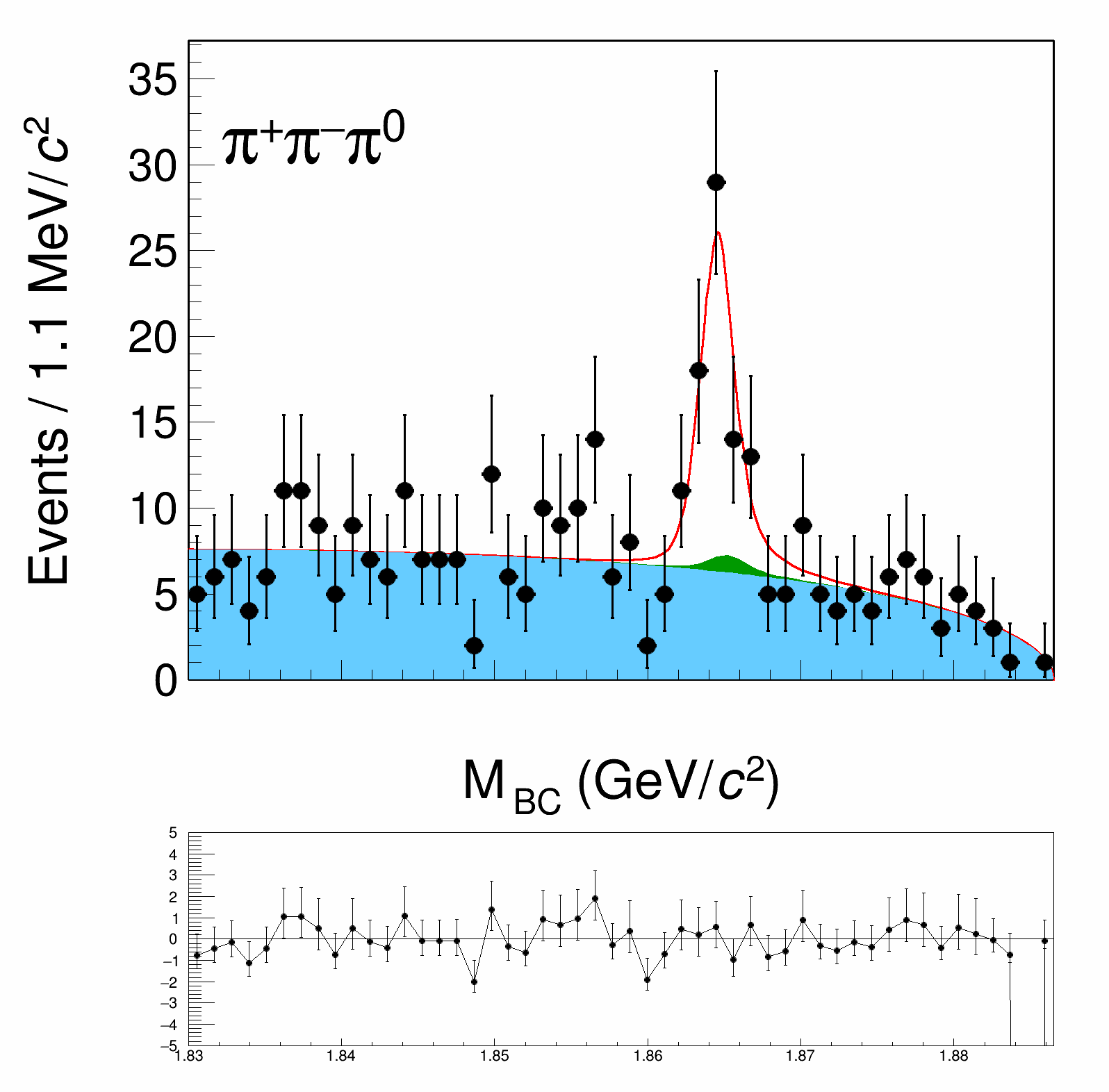}
    \includegraphics[height=3.6cm,trim={5.5cm 20.0cm 2.5cm 1.5cm},clip]{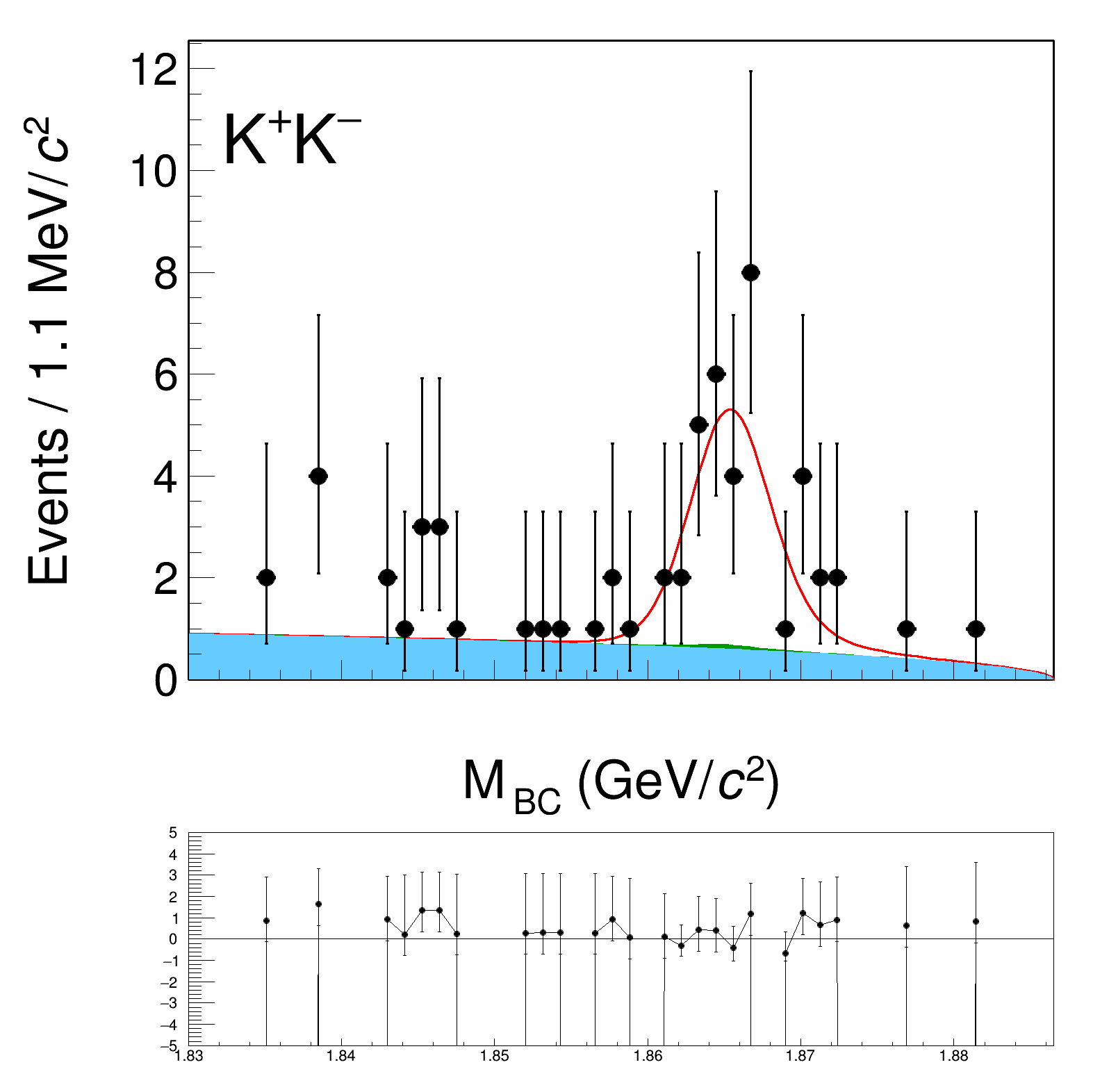}
    \includegraphics[height=3.6cm,trim={5.5cm 20.0cm 2.5cm 1.5cm},clip]{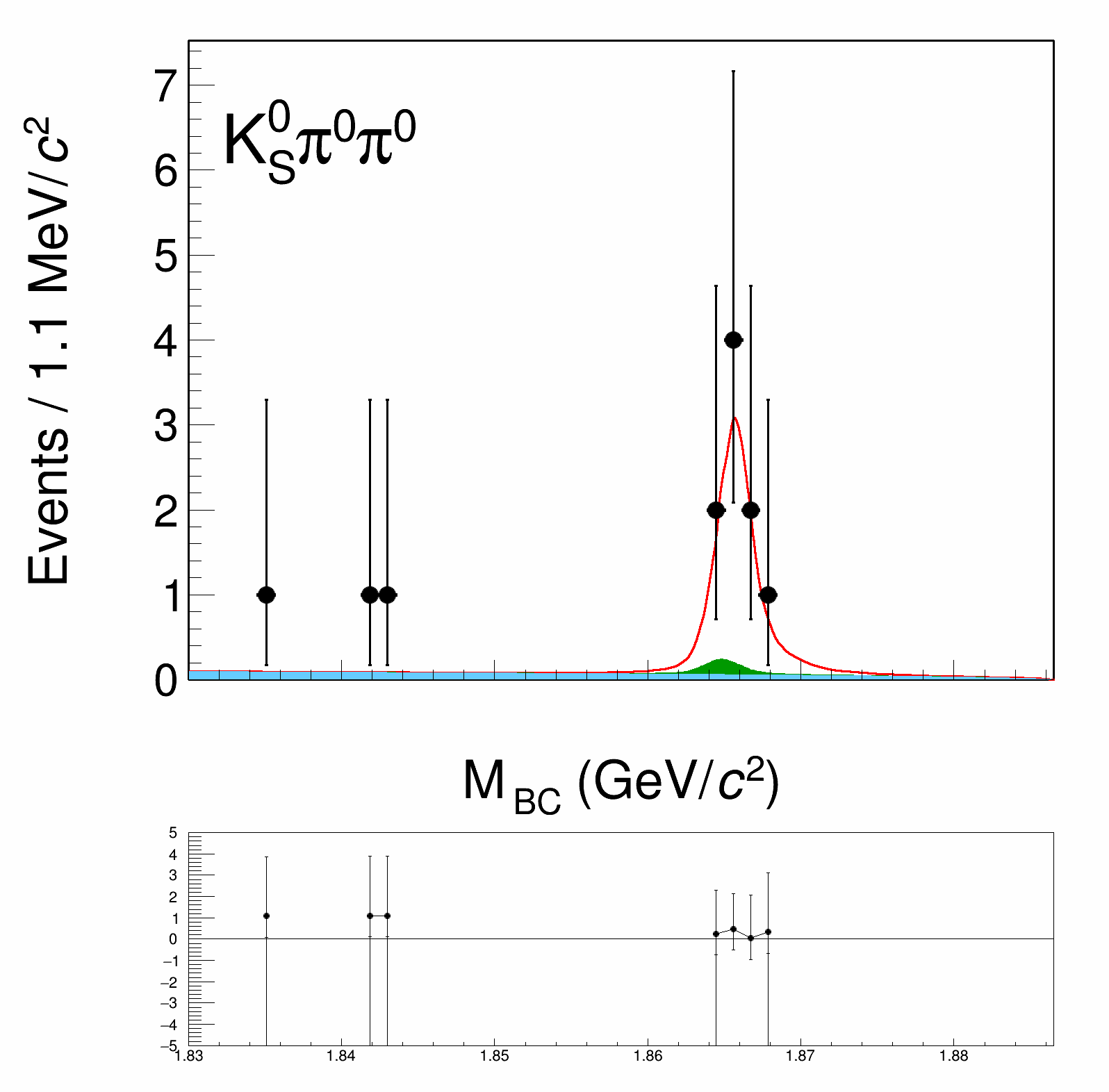}
    \includegraphics[height=3.6cm,trim={1.0cm 20.0cm 2.5cm 1.5cm},clip]{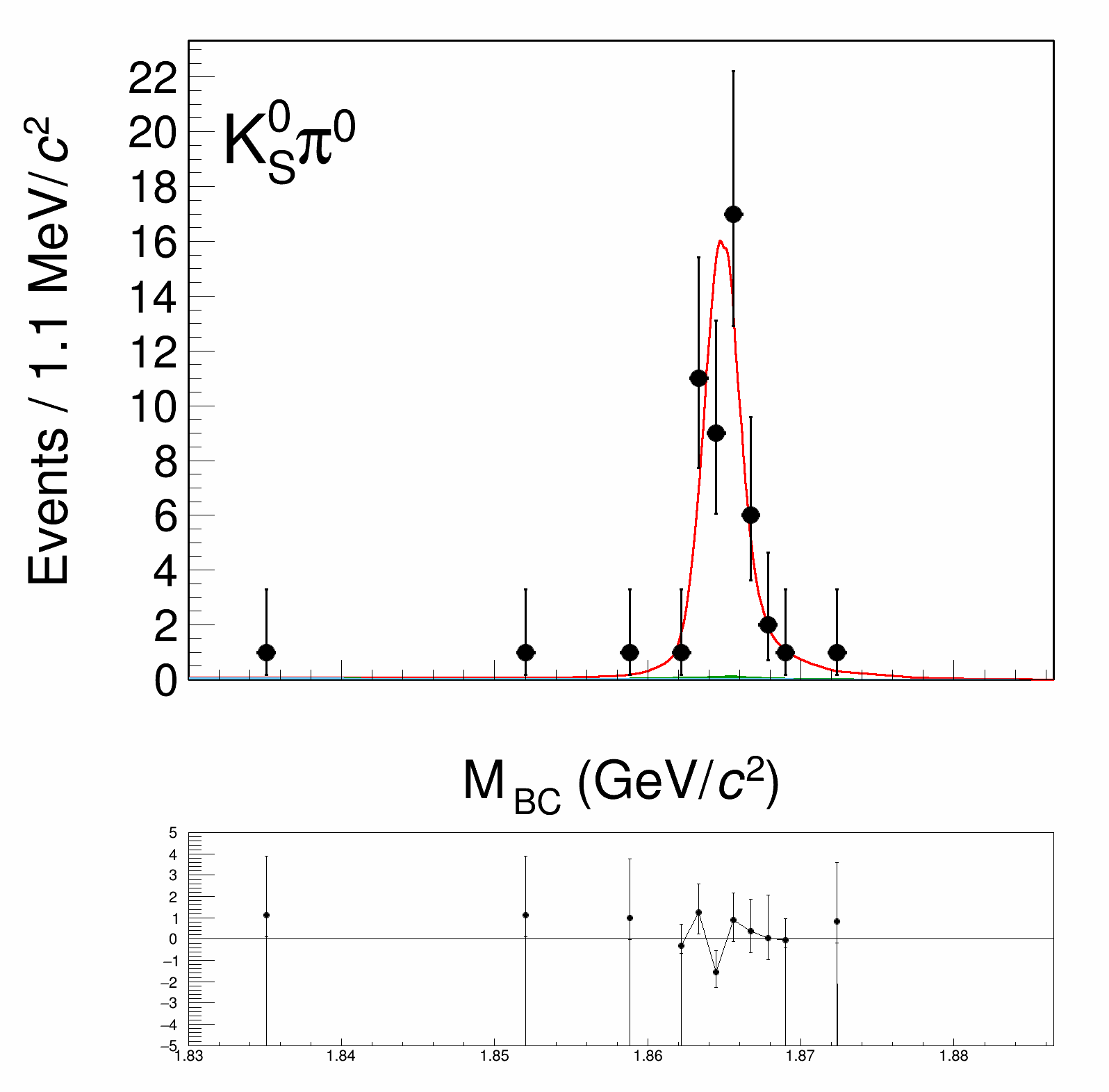}
    \includegraphics[height=3.6cm,trim={5.5cm 20.0cm 2.5cm 1.5cm},clip]{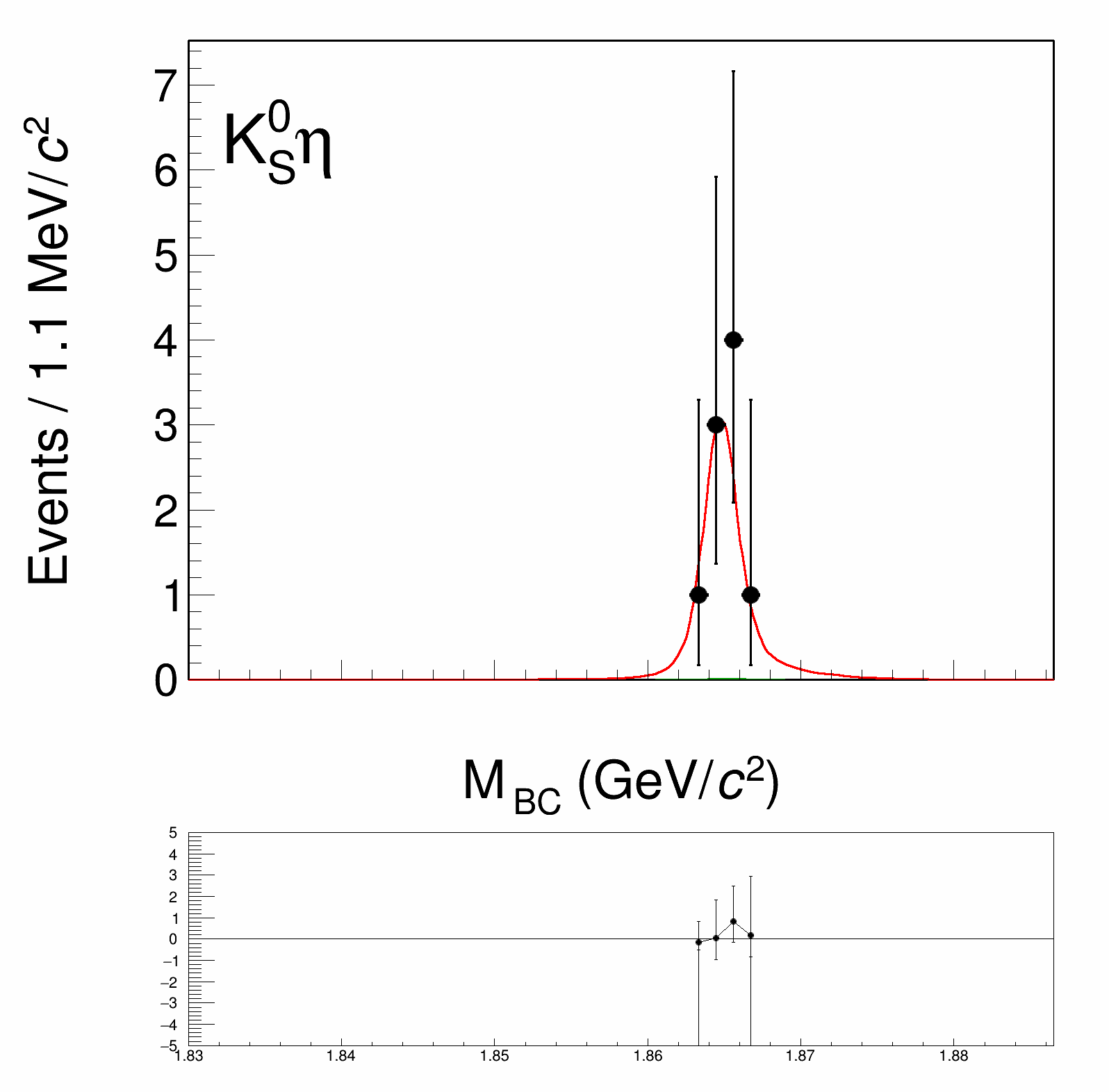}
    \includegraphics[height=3.6cm,trim={5.5cm 20.0cm 2.5cm 1.5cm},clip]{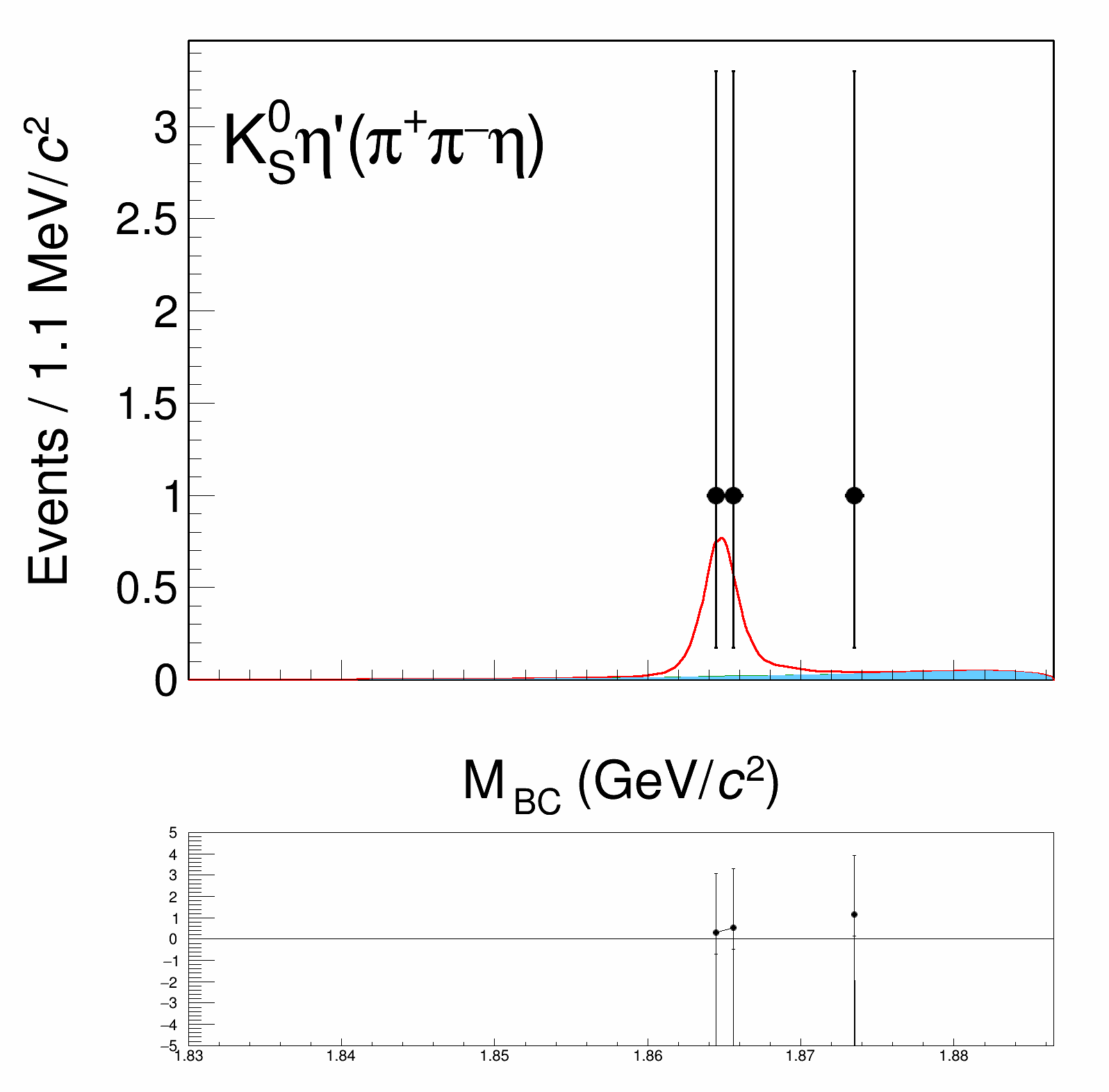}
    \includegraphics[height=4.3cm,trim={1.0cm 13.5cm 2.5cm 1.5cm},clip]{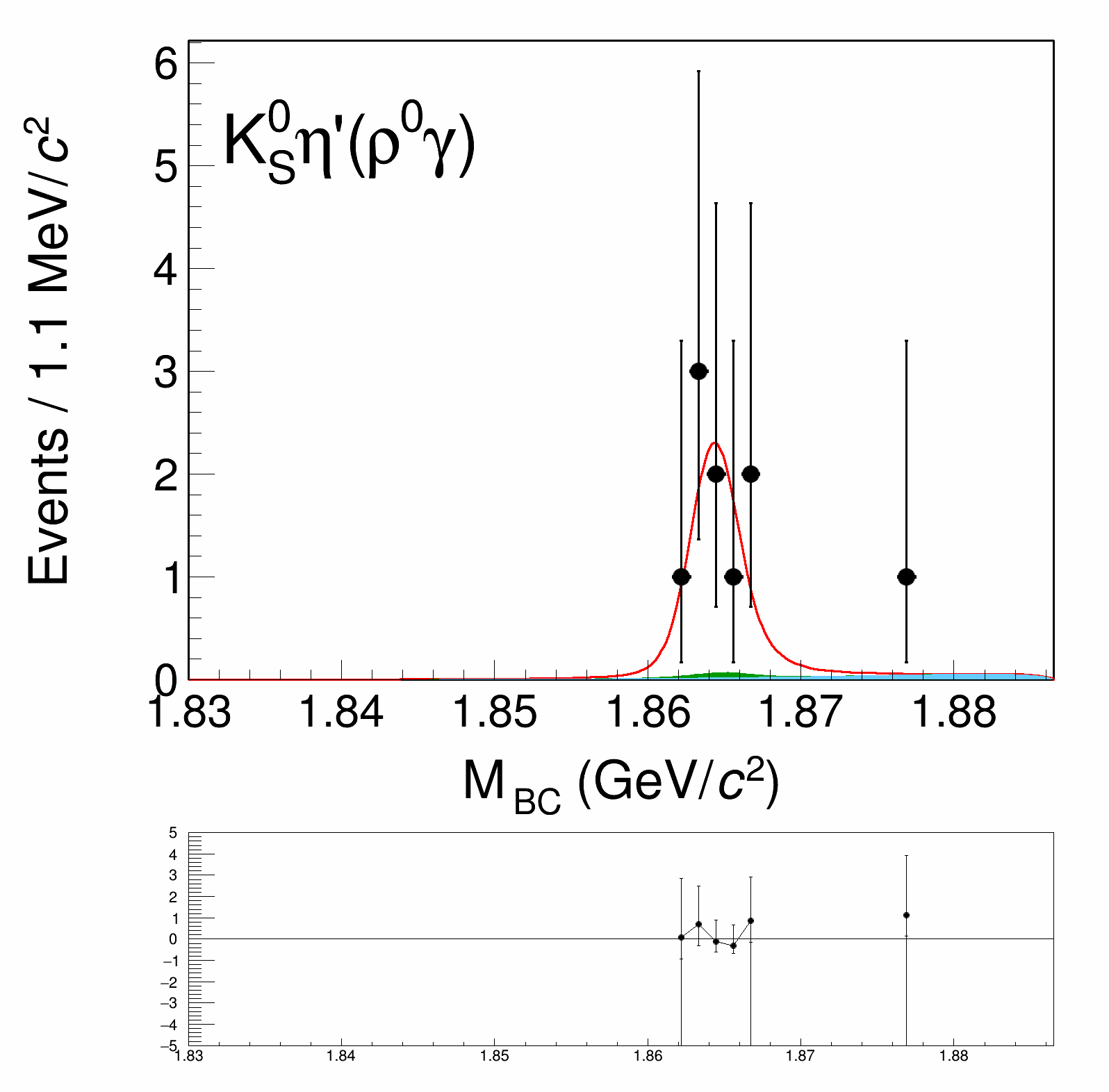}
    \includegraphics[height=4.3cm,trim={5.5cm 13.5cm 2.5cm 1.5cm},clip]{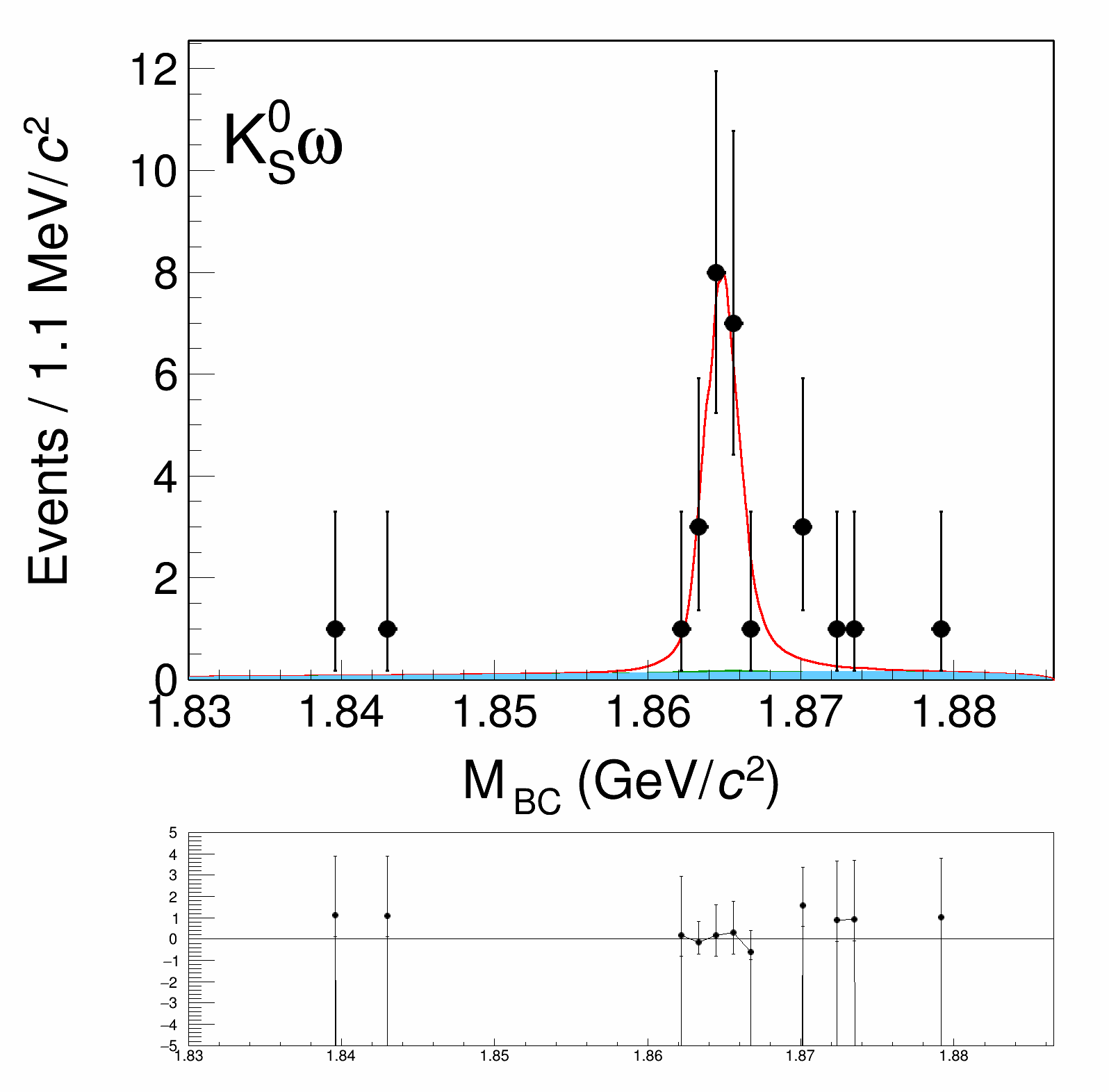}
    \includegraphics[height=4.3cm,trim={5.5cm 13.5cm 2.5cm 1.5cm},clip]{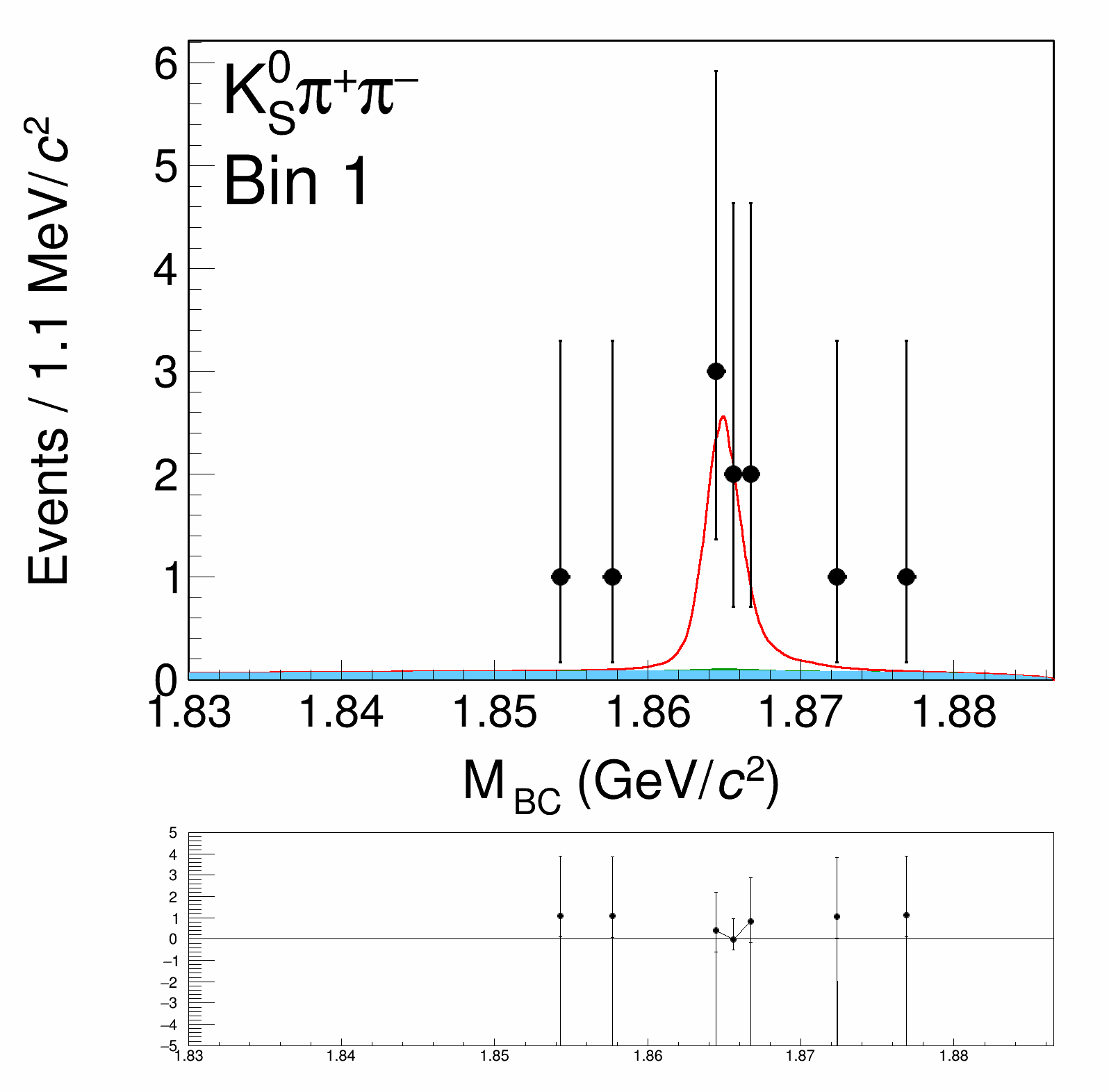}
    \caption{DT $M_{\rm BC}$ distributions of fully reconstructed DT candidates. Data points are shown in black with error bars and the red curve is the fit result. The solid blue shape is combinatorial background. The green, stacked on top of the blue, is peaking background.}
    \label{figure:DT_MBC}
\end{figure*}

\begin{figure*}[htb]
    \centering
    \includegraphics[height=4.1cm,trim={1.0cm 13.5cm 2.5cm 1.5cm},clip]{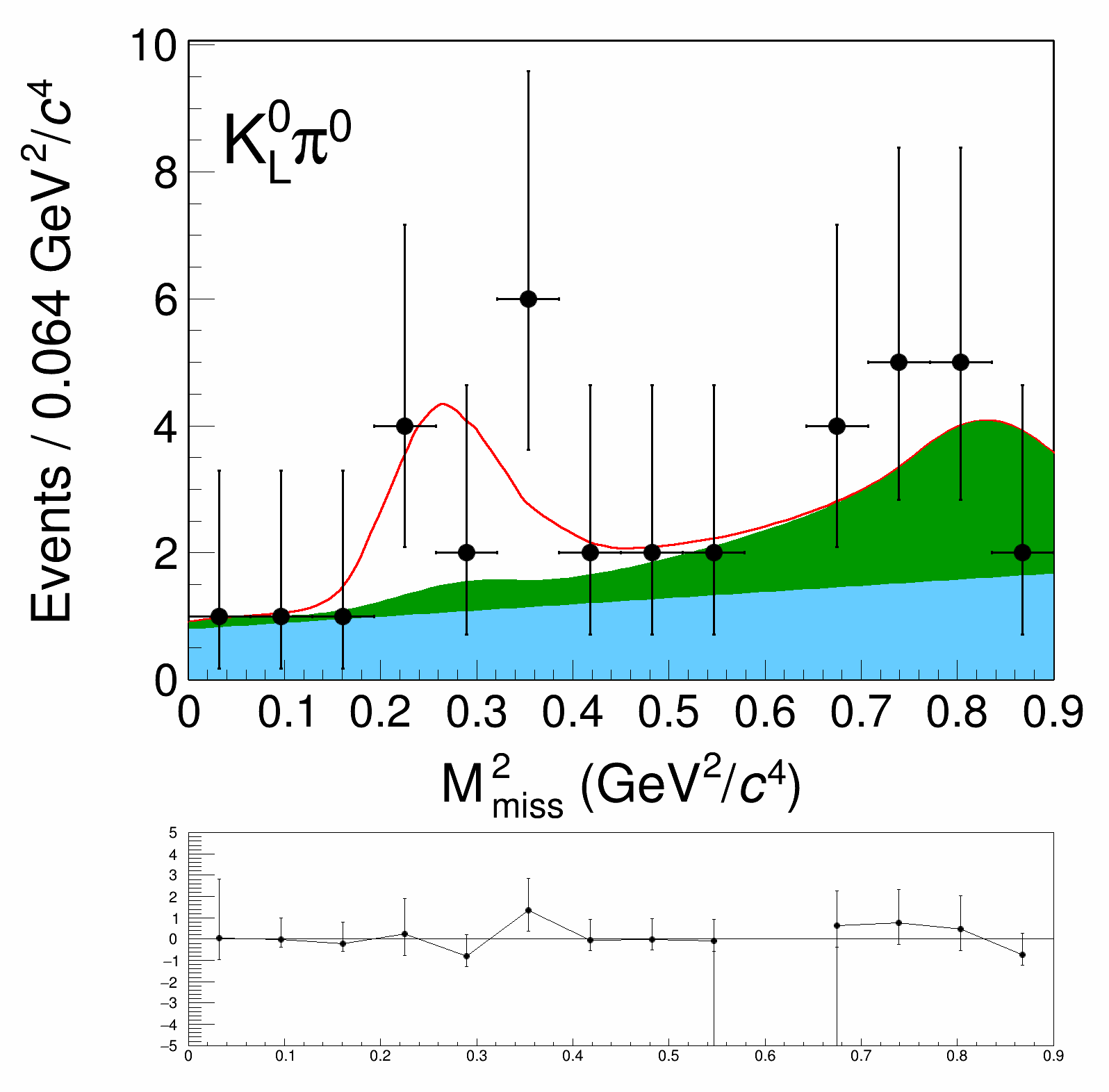}
    \includegraphics[height=4.1cm,trim={1.0cm 13.5cm 2.5cm 1.5cm},clip]{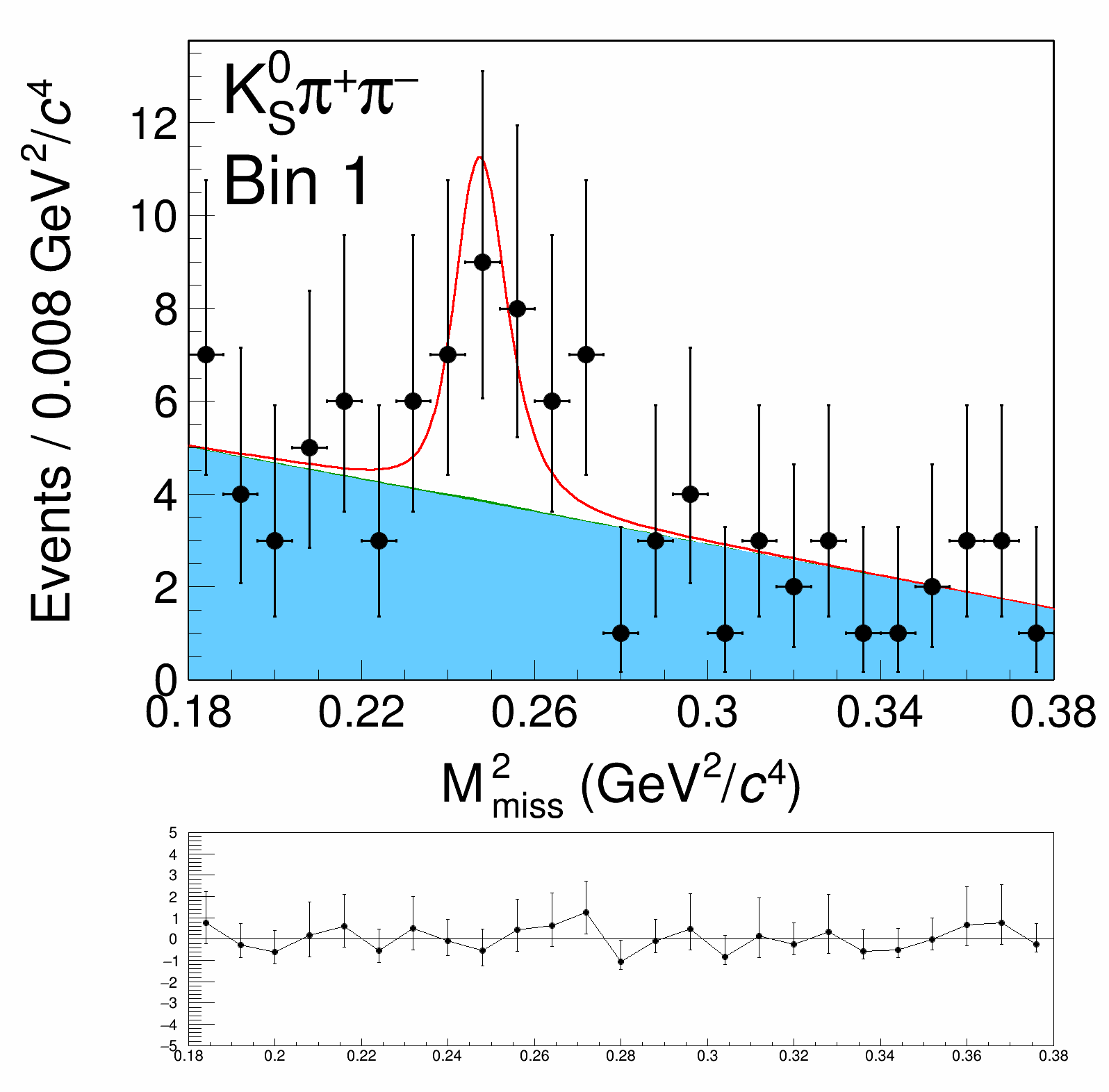}
    \includegraphics[height=4.1cm,trim={1.0cm 13.5cm 2.5cm 1.5cm},clip]{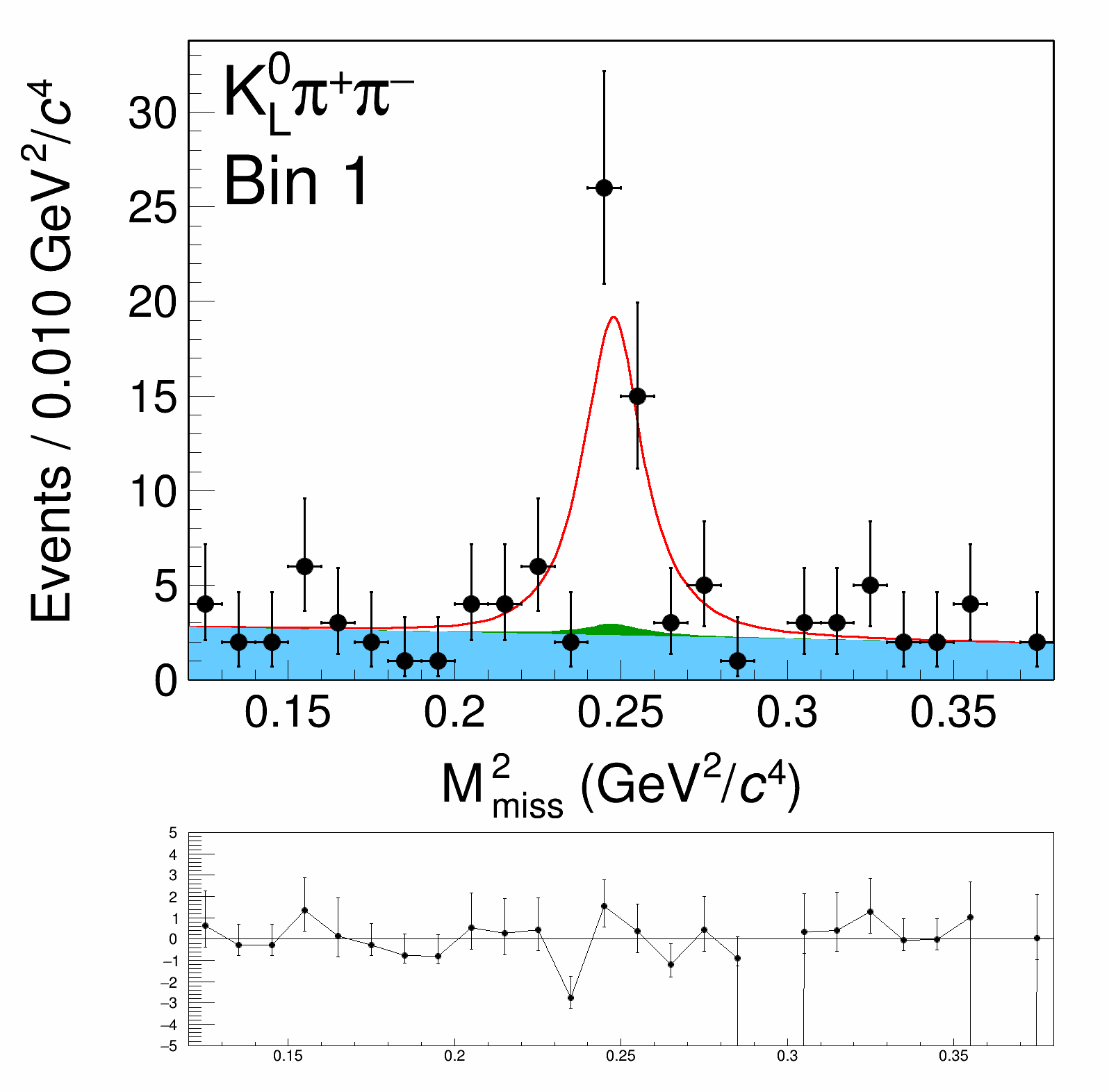}
    \caption{DT $M_{\rm miss}^2$ distributions of partially reconstructed DT candidates. Data points are shown in black with error bars and the red curve is the fit result. The solid blue shape is combinatorial background. The green, stacked on top of the blue, is peaking background.}
    \label{figure:DT_MMiss2}
\end{figure*}

For partially reconstructed tag modes, a fit of the missing-mass squared $M_{\rm miss}^2$ of the missing particle is performed instead. In the $K_{S, L}^0\pi^+\pi^-$ tag modes, which are split into bins of phase space, a simultaneous fit is performed where the signal-shape parameters are shared between all bins, while the yield of signal and combinatorial backgrounds are varied independently in each bin. Figure~\ref{figure:DT_MBC} shows the signal $M_{\rm BC}$ distributions of fully reconstructed modes in data, along with the fitted shapes. The corresponding $M_{\rm miss}^2$ distributions for partially reconstructed modes are shown in Fig.~\ref{figure:DT_MMiss2}. For the $K_{S, L}^0\pi^+\pi^-$ tags, only the result in one bin of phase space is shown, but the other bins are very similar. The fitted yields and their efficiencies, as determined from simulation, are listed in Table~\ref{table:Double_tag_yields_efficiencies_CP} and Table~\ref{table:Double_tag_yields_efficiencies_K0pipi}.

In the $D\to K^0_S\omega$ tag, there is also non-resonant $D\to K^0_S\pi^+\pi^-\pi^0$ background. To isolate $D\to K^0_S\omega$ candidates from non-resonant $D\to K^0_S\pi^+\pi^-\pi^0$, first the sPlot technique~\cite{cite:sPlot} is used on the $M_{\rm BC}$ variable to remove the flat combinatorial background. Then a fit of the $\pi^+\pi^-\pi^0$ invariant mass, after applying sWeights, is performed to obtain the yield of $K^0_S\omega$. This procedure is done with both ST and DT candidates.

\begin{table}[htb]
    \centering
    \caption{DT  yields and efficiencies for $C\!P$ tags. The uncertainties are statistical only.}
    \vspace{0.2cm}
    \label{table:Double_tag_yields_efficiencies_CP}
    \begin{tabular}{lcc}
        \hline
        Tag mode                            & Yield                & Efficiency ($\%$) \\
        \hline
        $K_S^0\omega$                       & $9 \pm 3$            & $2.23 \pm 0.02$    \\
        $K_S^0\eta^\prime(\pi^+\pi^-\eta)$  & $2 \pm 2$            & $2.02 \pm 0.02$    \\
        $K_S^0\eta^\prime(\rho^0\gamma)$    & $9 \pm 3$            & $3.29 \pm 0.03$    \\
        $K_S^0\eta$                         & $9 \pm 3$            & $5.72 \pm 0.04$      \\
        $K_S^0\pi^0$                        & $48 \pm 7$\phantom{0}           & $6.86 \pm 0.04$      \\
        $\pi^+\pi^-\pi^0$                   & $53 \pm 10$          & $7.67 \pm 0.04$      \\
        $\pi^+\pi^-$                        & $2 \pm 4$            & $15.07 \pm 0.06$\phantom{0}     \\
        $K_S^0\pi^0\pi^0$                   & $8 \pm 3$            & $2.87 \pm 0.03$    \\
        $K_L^0\pi^0$                        & $7 \pm 5$            & $5.30 \pm 0.04$      \\
        $K^+K^-$                            & $28 \pm 10$          & $14.55 \pm 0.06$\phantom{0}     \\
        \hline
    \end{tabular}
\end{table}

\begin{table}[htb]
    \centering
    \caption{DT  yields for $K_{S, L}^0\pi^+\pi^-$ tags, for full and partial reconstruction, in bins of phase space. The global efficiencies are also shown. The uncertainties are statistical only.}
    \vspace{0.2cm}
    \label{table:Double_tag_yields_efficiencies_K0pipi}
    \begin{tabular}{cccc}
        \hline
                                     & $K_S^0\pi^+\pi^-$            & $K_S^0\pi^+\pi^-$            & $K_L^0\pi^+\pi^-$            \\
        Bin                          & Full                         & Partial                      & Partial                      \\
        \hline
        1                            & $7 \pm 3$                    & $17 \pm 7$\phantom{0}                   & $45 \pm 8$                   \\
        2                            & $11 \pm 4$\phantom{0}                   & $4 \pm 4$                    & $15 \pm 5$                   \\
        3                            & $5 \pm 2$                    & $12 \pm 6$\phantom{0}                   & $19 \pm 5$                   \\
        4                            & $7 \pm 3$                    & $0 \pm 3$                    & \phantom{0}$5 \pm 3$                \\
        5                            & $11 \pm 3$\phantom{0}                   & $23 \pm 7$\phantom{0}                   & $11 \pm 5$                   \\
        6                            & $6 \pm 2$                    & $7 \pm 4$                    & $15 \pm 4$                   \\
        7                            & $12 \pm 3$\phantom{0}                   & $22 \pm 6$\phantom{0}                   & $22 \pm 5$                   \\
        8                            & $11 \pm 4$\phantom{0}                   & $6 \pm 5$                    & $25 \pm 6$                   \\
        \hline
        Total                        & $69 \pm 9$\phantom{0}                   & $91 \pm 15$                  & $158 \pm 15$\phantom{0}                 \\
        \hline
        $\epsilon_{\rm DT}$ ($\%$)   & \phantom{0}$6.56 \pm 0.04$\phantom{0}              & \phantom{0}$7.01 \pm 0.04\phantom{0}$              & \phantom{0}$7.25 \pm 0.04$\phantom{0}              \\
        \hline
    \end{tabular}
\end{table}

\section{\texorpdfstring{$C\!P$}{CP}-even fraction measurement}
\label{section:CP_even_fraction_measurement}

\begin{figure*}[htb]
    \centering
    \includegraphics[width=0.5\textwidth]{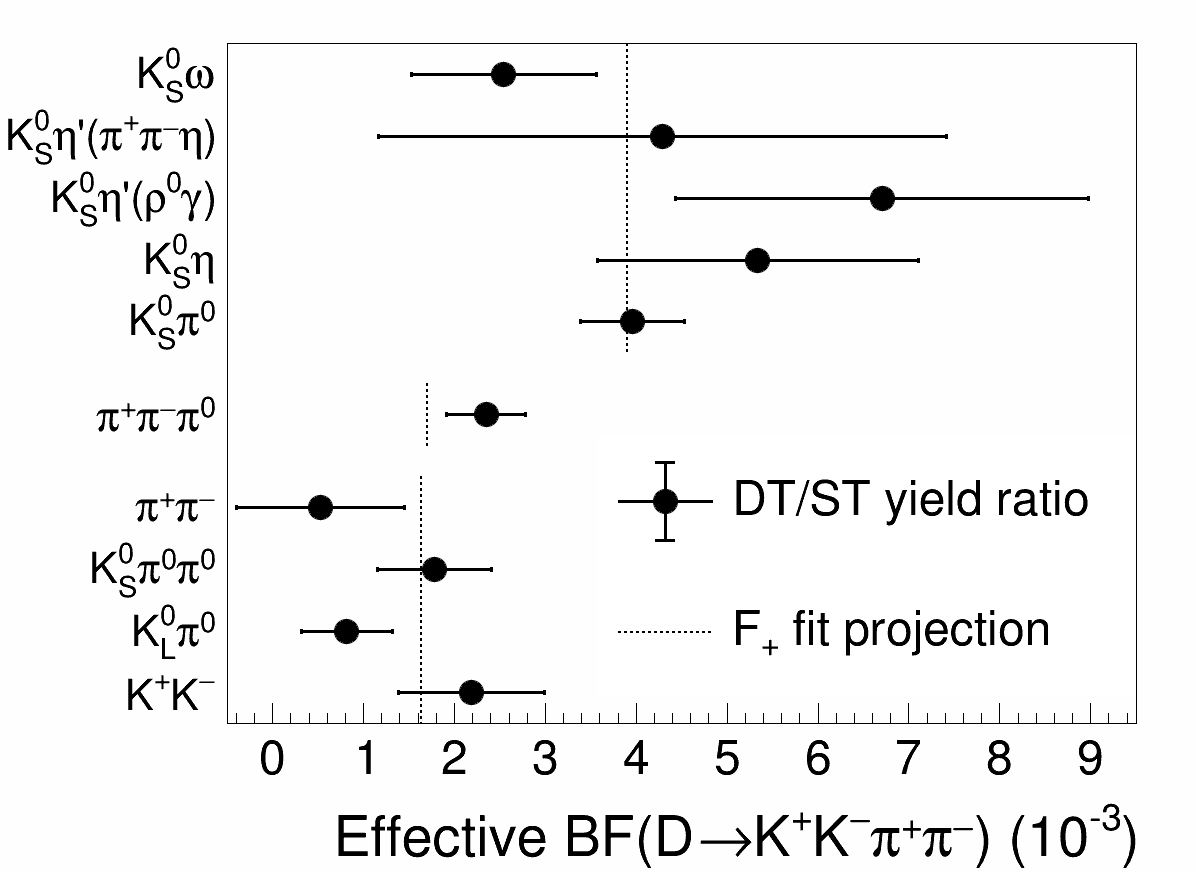}
    \caption{The effective branching fraction (BF) of $D\to K^+K^-\pi^+\pi^-$ measured against $C\!P$-odd (top), $D\to\pi^+\pi^-\pi^0$ and $C\!P$-even (bottom) tags. The black dotted lines indicate the values expected from the fit.}
    \label{figure:FPlus_CP_tags}
\end{figure*}

A maximum-likelihood fit is performed to the ST and DT yields of $C\!P$ tags listed in Table~\ref{table:Single_tag_yields_efficiencies} and Table~\ref{table:Double_tag_yields_efficiencies_CP}, respectively, assuming the relation given by Eq.~\eqref{equation:DT_ST_yield_ratio}. The uncertainties are assumed to follow a Gaussian distribution. The branching fraction $\mathcal{B}(KK\pi\pi)$ and the $C\!P$-even fraction $F_+$ are free parameters in the fit. Figure~\ref{figure:FPlus_CP_tags} shows the ratio of DT yields to ST yields for each tag after efficiency corrections. Physically, this represents the effective branching fraction of $D\to K^+K^-\pi^+\pi^-$, where the $D$ meson is prepared in a $C\!P$ eigenstate. The fitted $C\!P$-even fraction is $F_+ = 0.704 \pm 0.042 \pm 0.028$, where the first uncertainty is from the statistical uncertainties of ST and DT yields and the second uncertainty is the systematic uncertainty, discussed in Section~\ref{section:Systematic_uncertainties}. The obtained branching fraction is $\mathcal{B}(KK\pi\pi) = (2.8 \pm 0.3)\times 10^{-3}$, where the uncertainty is statistical. It is consistent with the current value from the Particle Data Group (PDG)~\cite{pdg}.

Similarly, a maximum-likelihood fit is performed using Eq.~\eqref{equation:DT_ST_yield_ratio_binned} and the $K_{S, L}^0\pi^+\pi^-$ results from Table~\ref{table:Double_tag_yields_efficiencies_K0pipi}. Table~\ref{table:Double_tag_yields_efficiencies_K0pipi} contains the global reconstruction efficiencies, but in the fit a full $8\times 8$ efficiency matrix is used to account for both the reconstruction efficiency in each bin, and the migration of events between the bins. The bin-migration effect is between $2$\%-$14\%$ for the fully reconstructed $K_S^0\pi^+\pi^-$ mode, $3$\%-$15\%$ for the partially reconstructed $K_S^0\pi^+\pi^-$ mode and $4$\%-$24\%$ for the $K_L^0\pi^+\pi^-$ mode.

Because the branching fraction of $D\to K_L^0\pi^+\pi^-$ is currently unknown, $\mathcal{B}(KK\pi\pi)$ is a free parameter and varied independently for the $K_S^0\pi^+\pi^-$ and $K_L^0\pi^+\pi^-$ tags. It is therefore not necessary to normalize the DT yield of the $D\to K_L^0\pi^+\pi^-$ tag with the corresponding ST yields and thus the measured $\mathcal{B}(KK\pi\pi)$ carry no useful information. For the $D\to K_S^0\pi^+\pi^-$ tag, the fitted branching fraction is $\mathcal{B}(KK\pi\pi) = (2.3 \pm 0.3)\times 10^{-3}$, where the uncertainty is statistical, which is compatible with both the known value and with the result from the $C\!P$ tags.

\begin{figure*}[htb]
    \centering
    \includegraphics[width=0.49\textwidth]{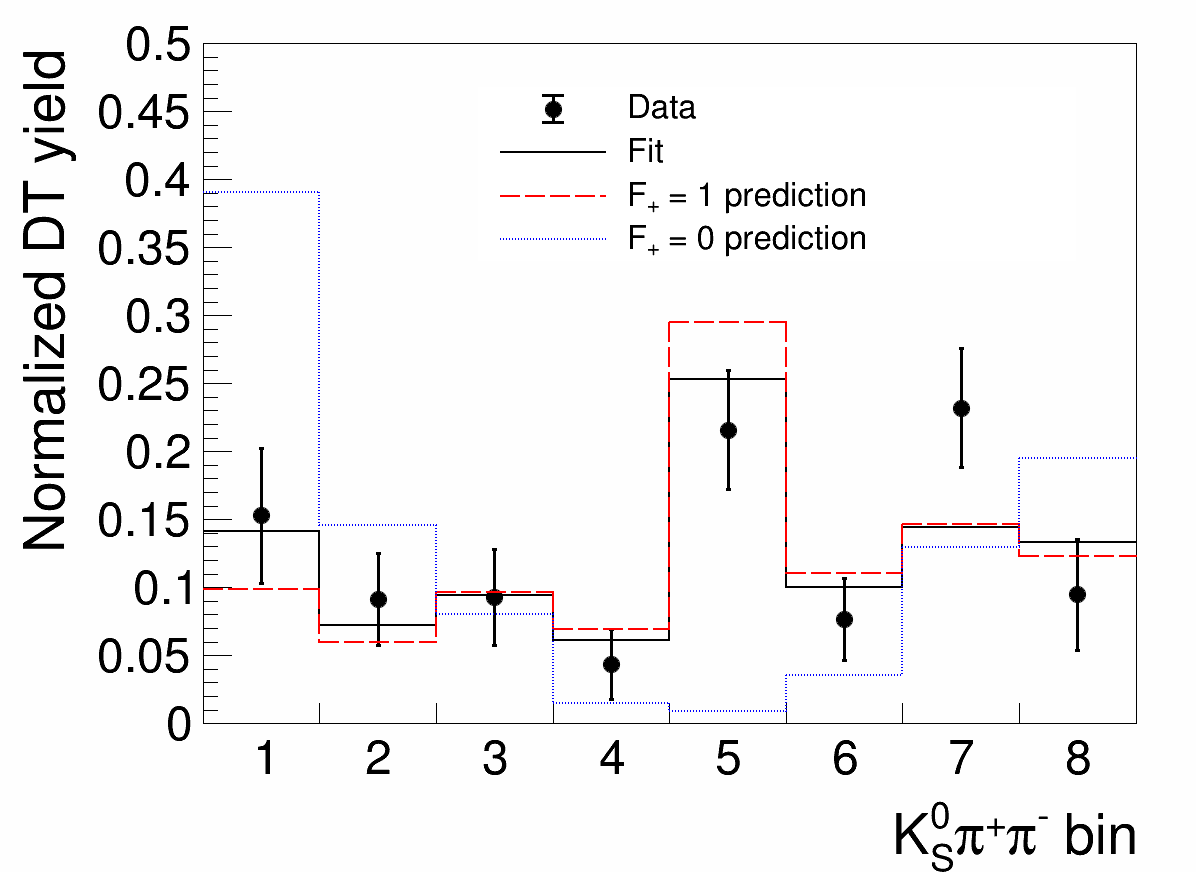}  
    \includegraphics[width=0.49\textwidth]{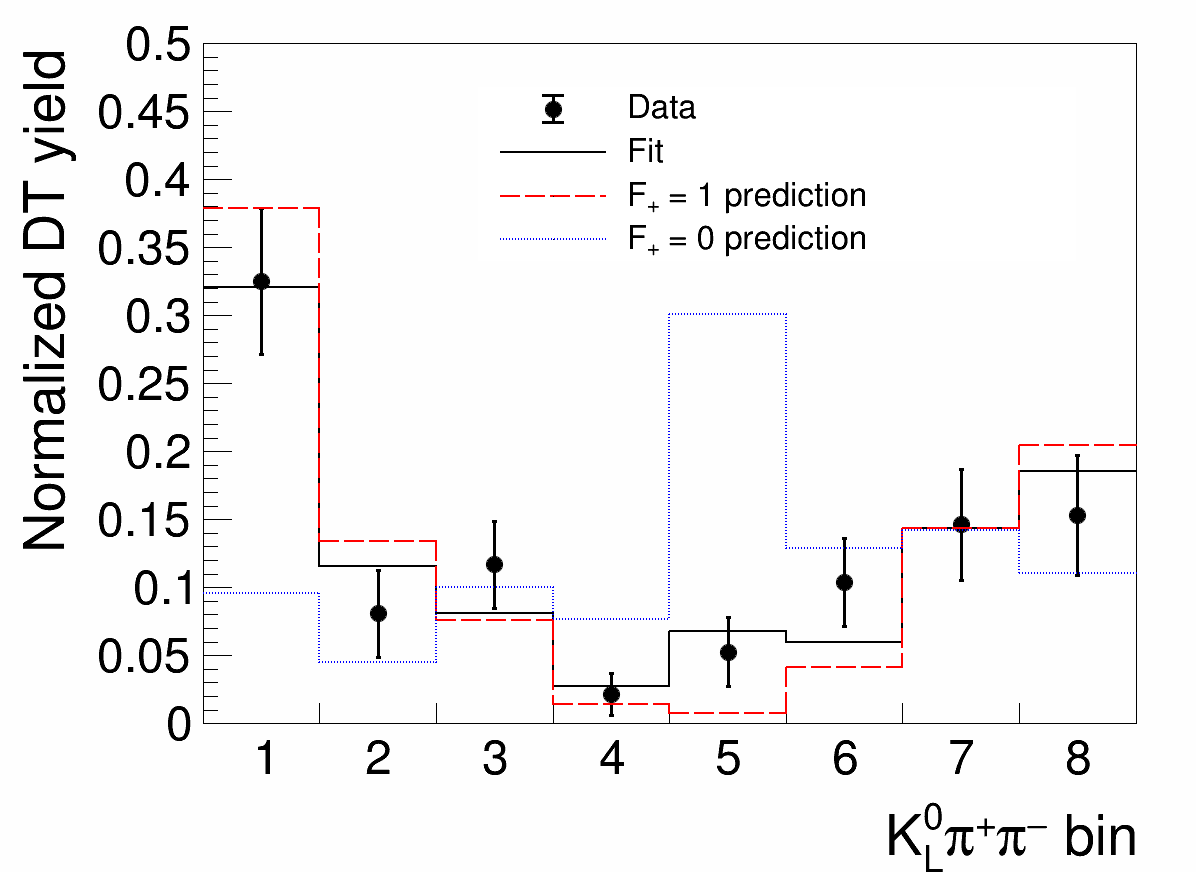}  
    \caption{Fit results for the $K^0_{S,L}\pi^+\pi^-$ tags.  Also shown is the fit projection, and the predictions for $F_+ = 1$ and $F_+ = 0$.}
    \label{figure:FPlus_K0pipi_tags}
\end{figure*}

Figure~\ref{figure:FPlus_K0pipi_tags} shows the DT yields in bins of phase space, where the sum of the yields has been normalized to unity. The $K_S^0\pi^+\pi^-$ plot contains both the fully reconstructed and partially reconstructed $K^+K^-\pi^+\pi^-$ vs $K_S^0\pi^+\pi^-$ data. The fit of $K_{S, L}^0\pi^+\pi^-$ results in $F_+ = 0.798 \pm 0.077 \pm 0.019$, which is consistent with the value obtained from $C\!P$ tags.

The combined measurement of the $C\!P$-even fraction, using both $C\!P$ tags and $K_{S, L}^0\pi^+\pi^-$ tags, and taking into account all correlations, is $F_+ = 0.730 \pm 0.037 \pm 0.021$. This result is in agreement with the central value $F_+ = 0.736$ predicted by the model of Ref.~\cite{LHCb-PAPER-2018-041}.

\section{Systematic uncertainties}
\label{section:Systematic_uncertainties}

\noindent Several sources of systematic uncertainties in the $F_+$ measurement are considered. The assigned values are given in Table~\ref{table:Systematic_uncertainties}, listed separately for the $C\!P$ tags and the $K_{S, L}^0\pi^+\pi^-$ tags.

Since the efficiencies are calculated from simulation samples of finite size, there are statistical uncertainties associated with their values. For the $C\!P$ tags, the efficiency corrections are single numbers, but in the case of the $K_{S, L}^0\pi^+\pi^-$ tags the efficiency corrections are matrices that also account for bin migration. To estimate the effect on $F_+$, the fit described in Sect.~\ref{section:CP_even_fraction_measurement} is repeated $1000$ times, each time smearing the efficiencies and efficiency matrices. The smearing of a parameter is performed by adding a random number, drawn from a Gaussian distribution with zero mean and a width equal to the uncertainty, to the parameter. The resulting width of the fitted $F_+$ values is taken as the systematic uncertainty.

The form of Eqs.~\eqref{equation:DT_ST_yield_ratio} and \eqref{equation:DT_ST_yield_ratio_binned} makes the analysis insensitive to biases arising from any imperfections in the modelling of the particle-reconstruction efficiencies in the MC simulation for the tag modes. Similarly, the determination of $F_+$ is robust against these same imperfections affecting the signal decay. The case of the $K_L^0\pi^0$ tag requires separate consideration. Here its effective ST yield has an uncertainty associated with the knowledge of the branching fraction and $N_{D\bar{D}}$. The value of the branching fraction that is input to the analysis derives from the measurement in Ref.~\cite{cite:deltaKpi}, which was performed on the same data set using a DT method. The systematic uncertainties associated with the $\pi^0$ reconstruction and the track veto in this measurement are common with the current analysis and hence cancel. The relative uncertainty on the branching fraction, with these contributions removed, is $3.3\%$. This uncertainty together with that on $N_{D\bar{D}}$, is propagated to the determination of $F_+$ by smearing the $K_L^0\pi^0$ effective ST yield. 

When the $C\!P$-tags yields are corrected for their efficiencies, it is implicitly assumed that the DT efficiencies factorize into a product of ST efficiencies in the same manner for all tags. The imperfections in this assumption are studied by repeating the determination of $F_+$ with all DT efficiencies replaced by a product of ST efficiencies. The resulting bias in $F_+$ is assigned as the systematic uncertainty arising from this factorization assumption.

In the fit of $C\!P$ tags, the $\pi^+\pi^-\pi^0$ tag mode requires an external input for its $C\!P$-even fraction $F_+^{\pi\pi\pi^0}$~\cite{cite:pipipi0_CPfraction}. Similarly, the fit of the $K_{S, L}^0\pi^+\pi^-$ tags requires external inputs for the $K_i$ and $c_i$ parameters~\cite{cite:CLEOcisiKSpipi, cite:KSpipiStrongPhase}. The systematic uncertainties arising from external inputs are estimated by smearing these parameters. The correlations are also accounted for in the smearing of the $c_i$ values.

In the determination of the ST and DT yields, there are systematic uncertainties arising from the peaking-background yields and the mass-shape parameterisation. It is found that the choice of parameterisation of the mass shapes has a negligible effect on the systematic uncertainties. When estimating the peaking-background contributions, the measured branching fractions of both signal and background have associated uncertainties~\cite{pdg} that must be accounted for. In addition, the quantum-correlation corrections also have uncertainties due to imperfect knowledge of the $C\!P$ contents. To propagate these to the measurement of $F_+$, the peaking-background yields are smeared in the fit described in Sec.~\ref{section:Single_and_double_tag_yield_determination} to first obtain a systematic uncertainty for the signal yields. Then the signal yields themselves are smeared in the $F_+$ fit to obtain the systematic uncertainty associated with the peaking backgrounds.

The $K_S^0$ veto removes $4\%$ of the $D\to K^+K^-\pi^+\pi^-$ phase space, and thus can perturb  $F_+$ from the value that corresponds to the inclusive decay. This potential bias is estimated by calculating $F_+$ using the model from Ref.~\cite{LHCb-PAPER-2018-041} with and without the veto. The difference is assigned as the systematic uncertainty, which is common to both $C\!P$ and $K_{S, L}^0\pi^+\pi^-$ tags.

\begin{table}[htb]
    \centering
    \caption{Summary of the sources of systematic uncertainty in the measurement of $F_+$, multiplied by $10^2$. Entries marked `/' indicate that the source is not relevant for the decay mode. The last two entries are fully correlated between the two classes of tag. } \vspace{0.2cm} 
    \label{table:Systematic_uncertainties}
    \begin{tabular}{lcccc}
        \hline
        Source                   & \hspace{0.1cm}$C\!P$ tags\hspace{0.1cm}   & \hspace{0.1cm}$K_{S, L}^0\pi^+\pi^-$ tags\hspace{0.1cm} \\
        \hline
        MC sample size           & $0.1$                    & $0.4$                   \\
        $K^0_L\pi^0$ ST yield    & $2.1$                    & /                     \\ 
        Efficiency factorisation & $0.6$                    & /                     \\ 
        External inputs          & $0.3$                    & $0.8$                   \\
        ST and DT yields         & $0.2$                    & $0.3$                   \\
        \hline
        $K_S^0$ veto             & $0.8$                    & $0.8$                   \\
        Efficiency reweighting   & $1.0$                    & $1.0$                   \\
        \hline
        Total                    & $2.6$                    & $1.6$                   \\
        \hline
    \end{tabular}
\end{table}

Finally, there is a systematic uncertainty due to the efficiency reweighting that accounts for any discrepancies between the data and the amplitude model. With the current precision, it can be assumed that this systematic uncertainty is common between all tags. This systematic uncertainty originates from Eqs.~\eqref{equation:DT_ST_yield_ratio} and \eqref{equation:DT_ST_yield_ratio_binned}, where it is seen that any imperfections in the modelling of the $D\to K^+K^-\pi^+\pi^-$ decay are not cancelled in the ratio, unlike the efficiency of the tag-side decay. The effect is studied with a data-driven strategy by using samples of ST $D\to K^+K^-\pi^+\pi^-$ candidates in data and simulation. Five invariant-mass variables are compared between data and simulation. Any discrepancies between data and simulation are removed by reweighting the simulation samples. Using these weights, the fit of $F_+$ is repeated and the change in the result is assigned as the systematic uncertainty. Since this systematic uncertainty calculation is data-driven, improved precision is also expected  with more data.

\section{Summary and outlook}
\label{section:Summary_and_conclusion}

\noindent The first model-independent measurement of the $C\!P$-even fraction $F_+$ of the decay mode $D^0 \to K^+K^-\pi^+\pi^-$ has been performed using ten $C\!P$-eigenstate tags and the self-conjugate multi-body modes $D\to K_{S, L}^0\pi^+\pi^-$, from a data sample of $e^+e^-\to\psi(3770)\to D\bar{D}$ events  corresponding to an integrated luminosity of $2.93$~fb$^{-1}$. The results combine to $F_+ = 0.730 \pm 0.037 \pm 0.021$, where the first uncertainty is statistical and the second is systematic, indicating that this decay mode has a high $C\!P$-even content. This result will be valuable for future measurements of the CKM-angle $\gamma$, and studies of charm mixing and $C\!P$ violation at LHCb and Belle II.

The measurement is dominated by statistical uncertainty and it will improve significantly with the larger charm-threshold data set that BESIII is expected to collect in the coming years~\cite{Ablikim:2019hff}. This increased sample size will also allow the study to be extended to localized regions of phase space, as has been done for other decay modes~\cite{cite:cisi4pi, cite:KSpipipi0, cite:KSpipiStrongPhase, cite:cisiKSKK, cite:K3piStrongPhase, cite:CLEOcisiKSpipi}.

\input{acknowledgement_2022-08-19}

\clearpage
\bibliographystyle{apsrev4-2}
\bibliography{References_no_arXiv}

\end{document}

%% file: authorlist_2022-08-19.tex
\author{\center BESIII Collaboration\\~\\
M.~Ablikim$^{1}$, M.~N.~Achasov$^{12,b}$, P.~Adlarson$^{72}$, M.~Albrecht$^{4}$, R.~Aliberti$^{33}$, A.~Amoroso$^{71A,71C}$, M.~R.~An$^{37}$, Q.~An$^{68,55}$, Y.~Bai$^{54}$, O.~Bakina$^{34}$
% ...
\vspace{0.2cm}
(BESIII Collaboration)\\
\vspace{0.2cm} {\it
$^{1}$ Institute of High Energy Physics, Beijing 100049, People's Republic of China\\
% ...
$^{4}$ Bochum  Ruhr-University, D-44780 Bochum, Germany\\
% ...
$^{12}$ G.I. Budker Institute of Nuclear Physics SB RAS (BINP), Novosibirsk 630090, Russia\\
% ...
$^{33}$ Johannes Gutenberg University of Mainz, Johann-Joachim-Becher-Weg 45, D-55099 Mainz, Germany\\
$^{34}$ Joint Institute for Nuclear Research, 141980 Dubna, Moscow region, Russia\\
% ...
$^{37}$ Liaoning Normal University, Dalian 116029, People's Republic of China\\
% ...
$^{54}$ Southeast University, Nanjing 211100, People's Republic of China\\
$^{55}$ State Key Laboratory of Particle Detection and Electronics, Beijing 100049, Hefei 230026, People's Republic of China\\
% ...
$^{68}$ University of Science and Technology of China, Hefei 230026, People's Republic of China\\
% ...
$^{71}$ University of Turin and INFN, (A)University of Turin, I-10125, Turin, Italy; (B)University of Eastern Piedmont, I-15121, Alessandria, Italy; (C)INFN, I-10125, Turin, Italy\\
$^{72}$ Uppsala University, Box 516, SE-75120 Uppsala, Sweden\\
% ...
$^{b}$ Also at the Novosibirsk State University, Novosibirsk, 630090, Russia\\
% ...
}
}

%% file: acknowledgement_2022-08-19.tex
\section*{Acknowledgement}

The BESIII collaboration thanks the staff of BEPCII and the IHEP computing center for their strong support. This work is supported in part by National Key R\&D Program of China under Contracts Nos. 2020YFA0406300, 2020YFA0406400; National Natural Science Foundation of China (NSFC) under Contracts Nos. 11635010, 11735014, 11835012, 11935015, 11935016, 11935018, 11961141012, 12022510, 12025502, 12035009, 12035013, 12192260, 12192261, 12192262, 12192263, 12192264, 12192265; the Chinese Academy of Sciences (CAS) Large-Scale Scientific Facility Program; Joint Large-Scale Scientific Facility Funds of the NSFC and CAS under Contract No. U1832207; the CAS Center for Excellence in Particle Physics (CCEPP); 100 Talents Program of CAS; The Institute of Nuclear and Particle Physics (INPAC) and Shanghai Key Laboratory for Particle Physics and Cosmology; ERC under Contract No. 758462; European Union's Horizon 2020 research and innovation programme under Marie Sklodowska-Curie grant agreement under Contract No. 894790; German Research Foundation DFG under Contracts Nos. 443159800, 455635585, Collaborative Research Center CRC 1044, FOR5327, GRK 2149; Istituto Nazionale di Fisica Nucleare, Italy; Ministry of Development of Turkey under Contract No. DPT2006K-120470; National Science and Technology fund; National Science Research and Innovation Fund (NSRF) via the Program Management Unit for Human Resources \& Institutional Development, Research and Innovation under Contract No. B16F640076; Olle Engkvist Foundation under Contract No. 200-0605; STFC (United Kingdom); Suranaree University of Technology (SUT), Thailand Science Research and Innovation (TSRI), and National Science Research and Innovation Fund (NSRF) under Contract No. 160355; The Royal Society, UK under Contracts Nos. DH140054, DH160214; The Swedish Research Council; U. S. Department of Energy under Contract No. DE-FG02-05ER41374.